\documentclass[preprint2]{aastex61}
\usepackage{txfonts}
\usepackage{rotating}

\usepackage{CJKutf8}

\shortauthors{Long et al.}

\begin{document}
\begin{CJK*}{UTF8}{gbsn}

\title{An ALMA Survey of faint disks in the Chamaeleon I star-forming region: Why are some Class II disks so faint?}  %

\author{Feng Long(龙凤)}
\affiliation{Kavli Institute for Astronomy and Astrophysics, Peking University, Beijing 100871, China}
\affiliation{Department of Astronomy, School of Physics, Peking University,  Beijing 100871, China}

\author{Gregory J. Herczeg(沈雷歌)}
\affiliation{Kavli Institute for Astronomy and Astrophysics, Peking University, Beijing 100871, China}

\author{Ilaria Pascucci}
\affiliation{Lunar and Planetary Laboratory, University of Arizona, Tucson, AZ 85721, USA}
\affiliation{Earths in Other Solar Systems Team, NASA Nexus for Exoplanet System Science}

\author{Daniel Apai}
\affiliation{Lunar and Planetary Laboratory, University of Arizona, Tucson, AZ 85721, USA}
\affiliation{Earths in Other Solar Systems Team, NASA Nexus for Exoplanet System Science}
\affiliation{Steward Observatory, University of Arizona, Tucson, AZ 85721, USA}

\author{Thomas Henning}
\affiliation{Max Planck Institute for Astronomy, Heidelberg, Germany}

\author{Carlo F. Manara}
\affiliation{European Southern Observatory, Karl-Schwarzschild-Str. 2, D-85748 Garching bei M\"{u}nchen, Germany}

\author{Gijs D. Mulders}
\affiliation{Lunar and Planetary Laboratory, University of Arizona, Tucson, AZ 85721, USA}
\affiliation{Earths in Other Solar Systems Team, NASA Nexus for Exoplanet System Science}

\author{L\'{a}szl\'{o}. Sz\H{u}cs}
\affiliation{Max-Planck-Institut f\"ur extraterrestrische Physik, Giessenbachstrasse 1, D-85748 Garching, Germany}

\author{Nathanial P. Hendler}
\affiliation{Lunar and Planetary Laboratory, University of Arizona, Tucson, AZ 85721, USA}


\begin{abstract}
ALMA surveys of nearby star-forming regions have shown that the dust mass in the disk is correlated with the stellar mass, but with a large scatter.  This scatter could indicate either different evolutionary paths of disks or different initial conditions within a single cluster. We present ALMA Cycle 3 follow-up observations for 14 Class II disks that were low S/N detections or non-detections in our Cycle 2 survey of the $\sim 2$ Myr-old Chamaeleon I star-forming region. With 5 times better sensitivity, we detect millimeter dust continuum emission from six more sources and increase the detection rate to 94\% (51/54) for Chamaeleon I disks around stars earlier than M3. The stellar-disk mass scaling relation reported in \citet{pascucci2016} is confirmed with these updated measurements. Faint outliers in the $F_{mm}$--$M_*$ plane include three non-detections (CHXR71, CHXR30A, and T54) with dust mass upper limits of 0.2 M$_\earth$ and three very faint disks (CHXR20, ISO91, and T51) with dust masses $\sim 0.5$ M$_\earth$. By investigating the SED morphology, accretion property and stellar multiplicity, we suggest for the three millimeter non-detections that tidal interaction by a close companion ($\la$100 AU) and internal photoevaporation may play a role in hastening the overall disk evolution. The presence of a disk around only the secondary star in  a binary system may explain the observed stellar SEDs and low disk masses for some systems. 

\end{abstract}


\section{Introduction} \label{sec:intro}
Protoplanetary disks orbiting young stars are the birthplaces of planetary systems. The lifetimes of disks set strong constraints on the timescale for giant planet formation. The disk frequency, as estimated from near- and mid-IR excess emission  and the presence of accretion, declines with age, leading to a typical disk lifetime of $\sim3-5$ Myr \citep[e.g.][]{haisch2001,hernandez2007,fedele2010}.  ALMA surveys of nearby star-forming regions show a declining trend of average dust disk mass with cluster age, for the set of stars that retain disks \citep[e.g.,][]{ansdell2016,ansdell2017,barenfeld2016,pascucci2016}.  Within any young cluster, the dust mass in the disk is correlated with the stellar mass, but with a large scatter.  Some disks are detected in the mid-IR but are undetected in the sub-mm.  
This scatter indicates that disks within a single cluster evolve differently, with a possible dependence on initial conditions \citep{mulders2017,lodato2017}, stellar multiplicity \citep[e.g.,][]{harris2012}, or other stellar or environmental properties. The transition from primordial to debris disks is still uncertain \citep{Moor2017}, but may be traced either by disks that are very faint in the sub-mm \citep[e.g.][]{hardy2015,espaillat2017} or by disks with inner cavities \citep[e.g.,][]{kim2009,merin2010,ercolano2017}.

Protoplanetary disks are physically and chemically complex systems that evolve through an interplay of different mechanisms. Current proposed disk evolution and/or dissipation mechanisms include viscous accretion \citep{hartmann1998}, photoevaporation \citep{clarke2001,owen2012,alexander2014}, dust grain growth and settling \citep{dullemond2005}, as well as dynamical sculpting by planets formed within \citep{lissauer2007,durisen2007,zhu2012,pinilla2015}. These mechanisms affect disk evolution around individual low-mass stars.  Since about half of stars are in multiple systems \citep[see review by][]{duchene2013}, stellar multiplicity adds another important factor altering the evolution of any circumstellar or circumbinary disk in a multiple system \citep{papaloizou1977,artymowicz1994,miranda2015,lubow2015}.

In a previous ALMA 887 $\mu$m survey, we characterized the scaling relation between dust disk mass ($M_{dust}$) and stellar mass ($M_\star$) for low-mass stars in the 2--3 Myr-old Chamaeleon I star-forming region \citep{pascucci2016}, located at an average distance of 190 pc\footnote{For this paper, we adopt an updated average distance for Chamaeleon I of $\sim$190 pc, based on an analysis of Gaia DR2 parallaxes \citep{gaia2018}.  This distance is consistent with the Gaia DR1 distance of 188 pc (\citealt{long2017}, see also \citealt{voirin2017}) but larger than the pre-Gaia distance of 160 pc adopted by \citet{pascucci2016}.}. 
The continuum emission was detected from $\sim$80\% of disks around stars with spectral type earlier than M3 (the \textit{Hot} sample) and $\sim$55\% of disks around stars with later spectral type (the \textit{Cool} sample).  The non-detections have low dust masses and may have undergone faster/different disk evolution or perhaps be in a transitional stage between primordial and debris disks.

In this paper, we present results from a follow-up ALMA survey of 887 $\mu$m continuum emission for the 14 faintest disks in the \textit{Hot} sample of \citet{pascucci2016}, including the 9 non-detections in the \textit{Hot} sample.  We describe this follow-up survey in Section \ref{sec:obs}, present results from this survey and describe the anomalously faint sources in Section \ref{sec:results} and \ref{sec:outliers}, and finally discuss the potential explanations for these faint sources as well as the implications of these results in Section \ref{sec:diss}.

\begin{deluxetable*}{lcccccc}[!t]
\tabletypesize{\scriptsize}
\tablecaption{ALMA Cycle 3 Observations\label{tab:ALMAObservations}}
\tablewidth{0pt}
\tablehead{
\colhead{UTC Date} & \colhead{Antenna} & \colhead{Baseline Range} & \colhead{pwv} & \multicolumn{3}{c}{Calibrators}\\ 
\colhead{ } & \colhead{Number} & \colhead{(m)} & \colhead{(mm)} & \colhead{Flux} &  \colhead{Bandpass}  &  \colhead{Phase}  
}
\startdata
2016 Jan 25 & 42 & 15-331 & 1.13 & J1107-4449 & J1427-4206 & J1058-8003 \\
2016 Mar 30 & 44 & 15-460 & 0.75 & J1107-4449 & J1427-4206 & J1058-8003 \\
\enddata
\tablecomments{All targets were observed on both nights.}
\end{deluxetable*}

\section{Observations and data reduction} \label{sec:obs}
In our ALMA Cycle 2 program 2013.1.00437 (PI: I. Pascucci), a sample of 93 protoplanetary disks (complete down to M6) in Chamaeleon I was observed in Band 7 \citep{pascucci2016}.  The disks were selected based on the presence of excess mid-IR emission and prior classification as a Class II object by \citet{luhman2008}.  The targets in this survey were split into a \textit{Hot} sample 
with shallow observations, and a \textit{Cool} sample 
with observations that were 5 times deeper. 
The block of stars in the \textit{Hot} sample was observed to a sensitivity of 1 mJy/beam and delivered prior to the Cycle 3 deadline, so we were able to request deeper observations of the faintest \textit{Hot} stars in our Cycle 3 program 2015.1.00333 (PI: I. Pascucci).  The name and spectral type for each of these sources are listed in Table \ref{tab:continuum}.

The ALMA Band 7 observational setup in Cycle 3 was similar to that in Cycle 2 (see the details of numbers of 12m antennas, baseline range, weather condition and calibrators listed in Table \ref{tab:ALMAObservations}). The three continuum basebands were centered at 333.8, 335.7, and 347.5 GHz with an aggregated bandwidth of 5.9 GHz and a weighting-averaged frequency of 340 GHz (882 $\mu$m)\footnote{The observed wavelength is slightly different from 887 $\mu$m in \citet{pascucci2016}, leading to a flux difference expected to be $<1$\%. In this paper, we use 887 $\mu$m when referring to the millimeter continuum band for consistency.}. The spectral line window was set to target $^{12}$CO $J = 3-2$, instead of the $^{13}$CO and C$^{18}$O $J = 3-2$ lines targeted in Cycle 2, because the CO isotopologue emission is very faint and usually undetected \citep{long2017}.  For each source, two sets of observations were executed for a total integration time of 3.6 min, reaching a 1$\sigma$ rms of 0.2 mJy beam$^{-1}$, compared to $\sim$1.0 mJy for the \textit{Hot} sample in our Cycle 2 data \citep{pascucci2016} .

\begin{figure*}[t]
\centering
    \includegraphics[width=0.95\textwidth, trim=0 500 60 0]{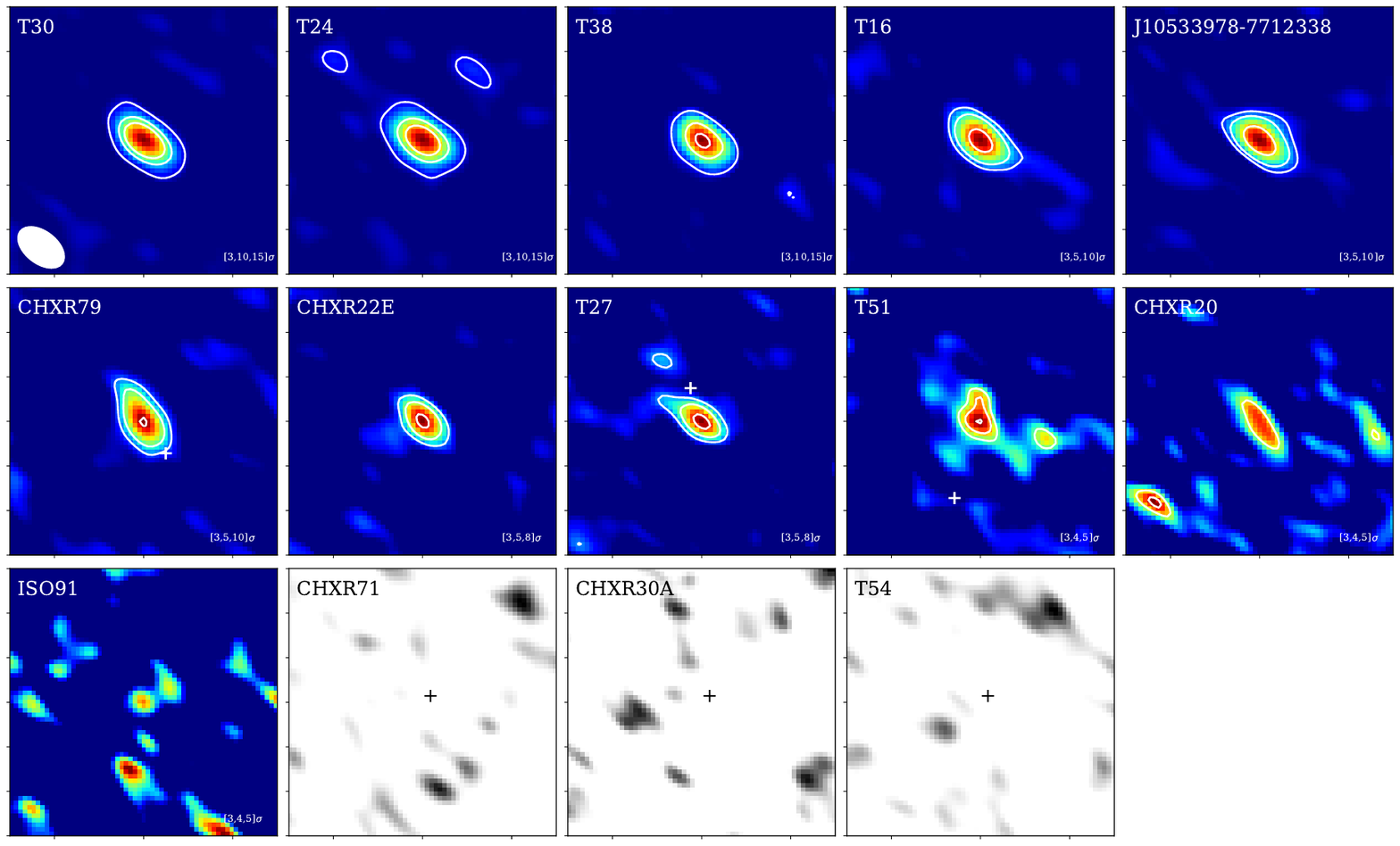} \\
    \caption{Continuum images at 887$\mu$m for Cycle 3 sample, ordered by decreasing millimeter flux with colormap scaled by the peak flux of each source. Contour levels for each panel are labeled at the right corner. Images are 6$\arcsec \times$ 6$\arcsec$ with the typical beam size of 1.$\arcsec$1$\times$0.$\arcsec$7 shown in the top left panel. Close companions for CHXR79, T27, and T51 are marked with white crosses, with position angles adopted from \citet{daemgen2013}. \label{fig:cont_images}}
\end{figure*}

The data calibration was performed using CASA 4.6.0, following the scripts provided by NRAO, including flux, phase, bandpass, and gain calibrations. We used J1427-4206 for bandpass calibration and J1058-8003 for gain calibration. We estimated the systematic flux uncertainty of $\sim$10$\%$ based on the amplitude and phase variations of the calibrators over time. Continuum images were created using the calibrated visibilities by averaging all the continuum channels, with natural weighting in \textit{clean} (see the continuum images in Figure \ref{fig:cont_images}).  The typical continuum beam size is 1.$\arcsec$1$\times$0.$\arcsec$7, when all baselines are included. Self-calibration was not applied on these weak sources.

\section{Results} \label{sec:results}
The continuum images for the 14 sources are shown in Figure \ref{fig:cont_images}. To measure continuum flux densities, we fit the visibility data using the CASA task \textit{uvmodelfit}, following the description in \citet{barenfeld2016}. This model has three free parameters: the integrated flux density ($F_\nu$), the phase center offsets in right ascension ($\Delta\alpha$) and in declination ($\Delta\delta$).  Most sources are unresolved and are well-fit with a point-source model.  The point-source models provide a good fit for low signal-to-noise detections and for sources with small aspect ratios.  
As in our Cycle 2 survey, elliptical models are also applied in \textit{uvmodelfit} to all detected sources with a signal-to-noise larger than 5, with three additional parameters: the FWHM of the major axis, the aspect ratio, and the position angle.  The elliptical Gaussian profiles provide a better fit and are adopted for only two sources, T16 and T24, for which lower flux uncertainties are obtained\footnote{CHXR20 looks like an edge-on disk, but the elliptical Gaussian model from \textit{uvmodelfit} returns a highly uncertain minor axis and \textit{imfit} do not yield a deconvolved source size because of the low signal-to-noise.}.
The uncertainties of the fitted parameters are scaled by the square root of the reduced $\chi^{2}$ of the fit.  The fitted fluxes and phase center offsets are consistent within the uncertainties for both the elliptical Gaussian and the point-source models.


\begin{deluxetable*}{ccccccccccr}
\tablecaption{Source Properties and Measured Fluxes \label{tab:continuum}}
\tablewidth{0pt}
\tablehead{
\colhead{2MASS} & \colhead{Name} & \colhead{Multiplicity} & \colhead{SpTy}  &\colhead{log(M$_{\ast}$)} & \colhead{F$_{mm}$(P16)} & \colhead{F$_{mm}$} & \colhead{$\Delta \alpha$} &\colhead{$\Delta \delta$} & \colhead{FWHM} & \colhead{log(M$_{dust}$)}\\
\colhead{ } &\colhead{ } &  \colhead{(arcsec)} &\colhead{ } &\colhead{(M$_{\sun}$)} & \colhead{(mJy)} & \colhead{(mJy)} & \colhead{(arcsec)} &  \colhead{(arcsec)} &  \colhead{(arcsec)} & \colhead{(M$_{\earth}$)} } 
\colnumbers
\startdata
J10533978-7712338 &         &        &   M2 & -0.41 &  4.60$\pm$0.79 &   3.74$\pm$0.16 & -0.47$\pm$0.02 & -0.04$\pm$0.01 &                   &  0.24 \\
J11023265-7729129 &  CHXR71 &  0.56  &   M3 & -0.52 & -0.21$\pm$0.82 &  -0.00$\pm$0.16 &       ...      &        ...     &                   & $<$-0.65 \\
J11045701-7715569 &     T16 &        &   M3 & -0.53 &  2.54$\pm$0.81 &   3.74$\pm$0.26 & -0.34$\pm$0.02 & -0.05$\pm$0.02 &  0.47$\times$0.22 &  0.24 \\
J11064510-7727023 &  CHXR20 & 28.46  &   K6 & -0.03 &  0.53$\pm$0.82 &   0.93$\pm$0.16 & -0.12$\pm$0.06 &  0.02$\pm$0.06 &                   & -0.36 \\
J11070925-7718471 &  ISO91 &        &   M3 & -0.52 &  0.06$\pm$0.82 &   0.65$\pm$0.17 &  0.34$\pm$0.09 & -0.03$\pm$0.09 &                   & -0.51 \\
J11071206-7632232 &     T24 &        &   M0 & -0.23 &  4.23$\pm$0.81 &   6.57$\pm$0.25 & -0.44$\pm$0.01 & -0.04$\pm$0.01 &  0.33$\times$0.11 &  0.49 \\
J11071330-7743498 & CHXR22E &        &   M4 & -0.63 &  0.42$\pm$0.81 &   2.23$\pm$0.16 & -0.64$\pm$0.03 &  0.08$\pm$0.02 &                   & 0.02 \\
J11072825-7652118 &     T27 &  0.78  &   M3 & -0.53 &  1.50$\pm$0.81 &   2.20$\pm$0.16 & -0.15$\pm$0.03 & -0.02$\pm$0.02 &                   & 0.01 \\
J11075809-7742413 &     T30 &        &   M3 & -0.51 &  6.45$\pm$0.79 &   7.11$\pm$0.16 & -0.37$\pm$0.01 &  0.00$\pm$0.01 &                   &  0.52 \\
J11080002-7717304 & CHXR30A &  0.46  &   K7 & -0.18 & -0.69$\pm$0.80 &  -0.07$\pm$0.15 &        ...     &       ...      &                   & $<$-0.67 \\
J11085464-7702129 &     T38 &        & M0.5 & -0.18 &  3.90$\pm$0.79 &   4.03$\pm$0.16 & -0.14$\pm$0.01 & -0.11$\pm$0.01 &                   &  0.28 \\
J11091812-7630292 &  CHXR79 &  0.88  &   M0 & -0.18 &  1.30$\pm$0.79 &   2.78$\pm$0.16 & -0.26$\pm$0.02 &  0.09$\pm$0.02 &                   &  0.12 \\
J11122441-7637064 &     T51 &  1.97  &   K2 &  0.04 &  0.19$\pm$0.78 &   1.21$\pm$0.16 & -0.39$\pm$0.05 & -0.12$\pm$0.05 &                   & -0.24 \\
J11124268-7722230 &    T54 &  0.24  &   K0 &  0.20 & -0.02$\pm$0.79 &  -0.00$\pm$0.16 &        ...     &        ...     &                   & $<$-0.65 \\
\enddata
\tablecomments{References for stellar multiplicity are \citet{lafrenire2008} and \citet{daemgen2013}. Spectral types are adopted from \citet{manara2016,manara2017} and stellar masses are adopted from \citet{pascucci2016}. Column 6 and 7 are 887$\mu$m fluxes in \citet{pascucci2016} and this work. Phasecenter offsets in R.A. and Dec are listed in column 8 and 9. When an FWHM value is listed in column 10, the source is fitted with an elliptical Gaussian. Otherwise, a point-source model is applied.  Dust masses in column 11 are calculated based on the same assumption as in \citet{pascucci2016} and for a constant temperature of 20 K. }

\end{deluxetable*}

\begin{figure}
    \includegraphics[width=0.45\textwidth, trim=0 0 0 0]{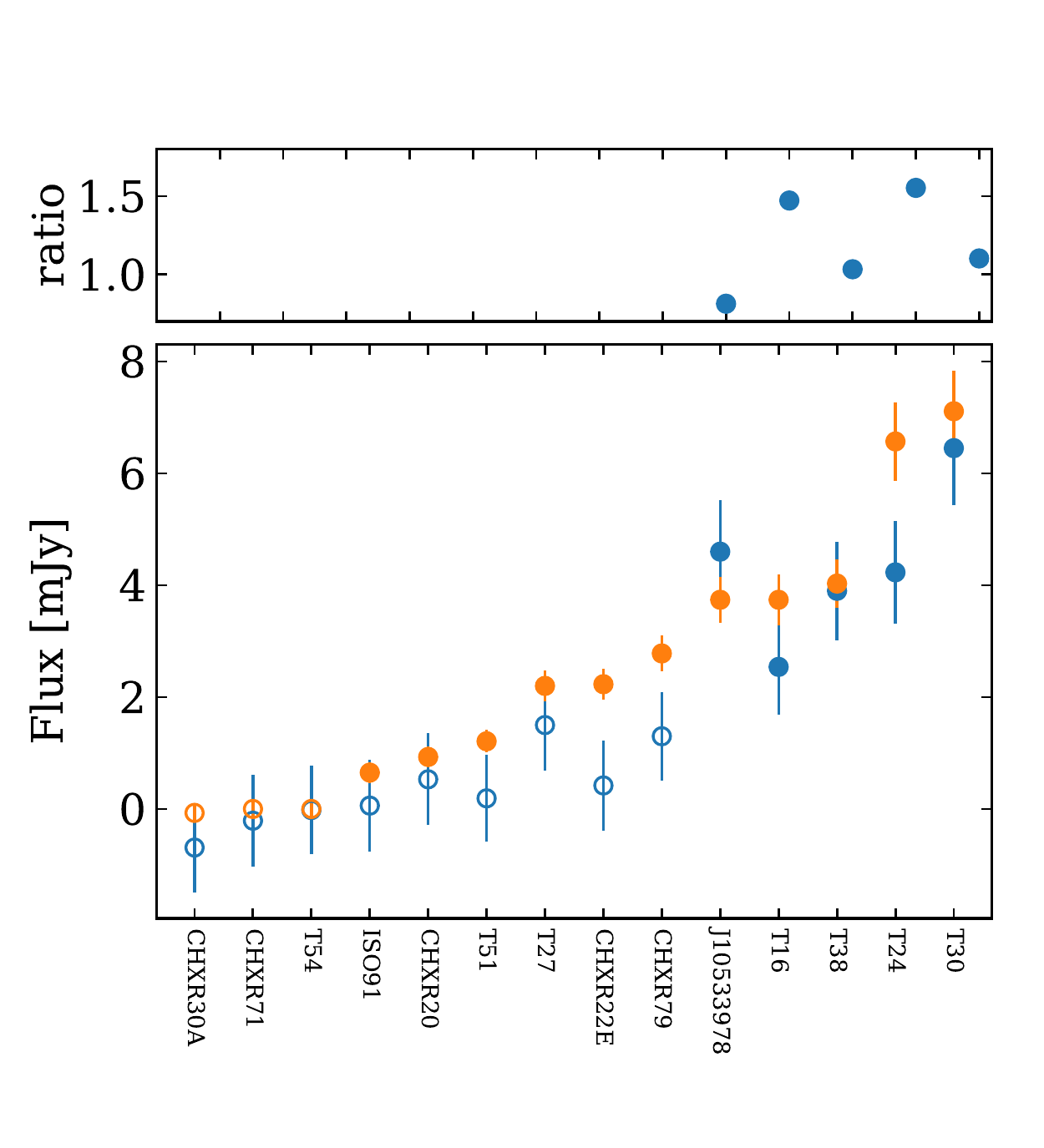}
    \caption{Bottom: Flux densities for Cycle 3 sources in order of increasing flux, with error bars including both the statistic uncertainty and the 10\% absolute flux calibration uncertainty (added in quadradure). Blue symbols are results from Cycle 2, while orange for Cycle 3. Solid dots and open circle represent detection and non-detection; Top: Flux ratios for the 5 sources that are detected in both observations. The average ratio is 1.2. \label{fig:flux}}
\end{figure}

Of the 14 observed sources in our Cycle 3 program, 11 are detected  with a signal-to-noise ratio of $>$3. Table \ref{tab:continuum} summarizes the measured continuum flux densities and associated uncertainties ($F_{mm}$), offsets from the phase center in right ascension and declination ($\Delta\alpha$ and $\Delta\delta$), and the source FWHMs for the two cases where an elliptical Gaussian model is adopted. Flux densities that are brighter than three times the uncertainty and located near the expected source position are considered detections. The median values of $\Delta\alpha$ = $-0.\arcsec34$ and $\Delta\delta$ = $-0.\arcsec02$ from the detections are consistent with the values reported in \citet{pascucci2016} in the full sample. These offsets from the 2MASS coordinates are roughly consistent with expectations based on proper motion \citep{lopez2013,murphy2013}.


Three sources remain undetected. For these sources (CHXR71, CHXR30A, and T54), the continuum emission is re-calculated by fitting a point-source model with fixed phase center offsets of $\Delta\alpha$ = $-0.\arcsec34$ and $\Delta\delta$ = $-0.\arcsec02$.  Table \ref{tab:continuum} lists the best-fit flux and uncertainty for these non-detections. These uncertainties do not include the 10\% absolute flux calibration error.

Six of the 11 detections here were undetected (at a 3$\sigma$ significance) in our Cycle 2 data.  The signal-to-noise ratios of the detections improved by a factor of 5.  Combining the observations presented in \citet{pascucci2016} and in this paper, the detection rate for the \textit{Hot} Sample is now 94\% (51/54).
The fluxes from Cycle 3 are generally higher than those observed in Cycle 2 (Figure \ref{fig:flux}), though consistent within 1-2$\sigma$ uncertainties. The flux ratios of the sources detected in both observations range from 0.8--1.5, with an average value of 1.2. We note that the flux calibrator, Pallas, used in our Cycle 2 observations has been reported recently to be problematic due to model uncertainties (Daniel Harsono, priv. communication; see also more description in CASA Task \textit{setjy}). In addition, the large phase variations in our shallow Cycle 2 observation affect the flux measurements for faint disks \citep{carilli1999}. The deeper Cycle 3 flux measurement is therefore more reliable than that of Cycle 2.

\begin{figure*}[t]
    \includegraphics[width=0.5\textwidth, trim=0 0 0 0]{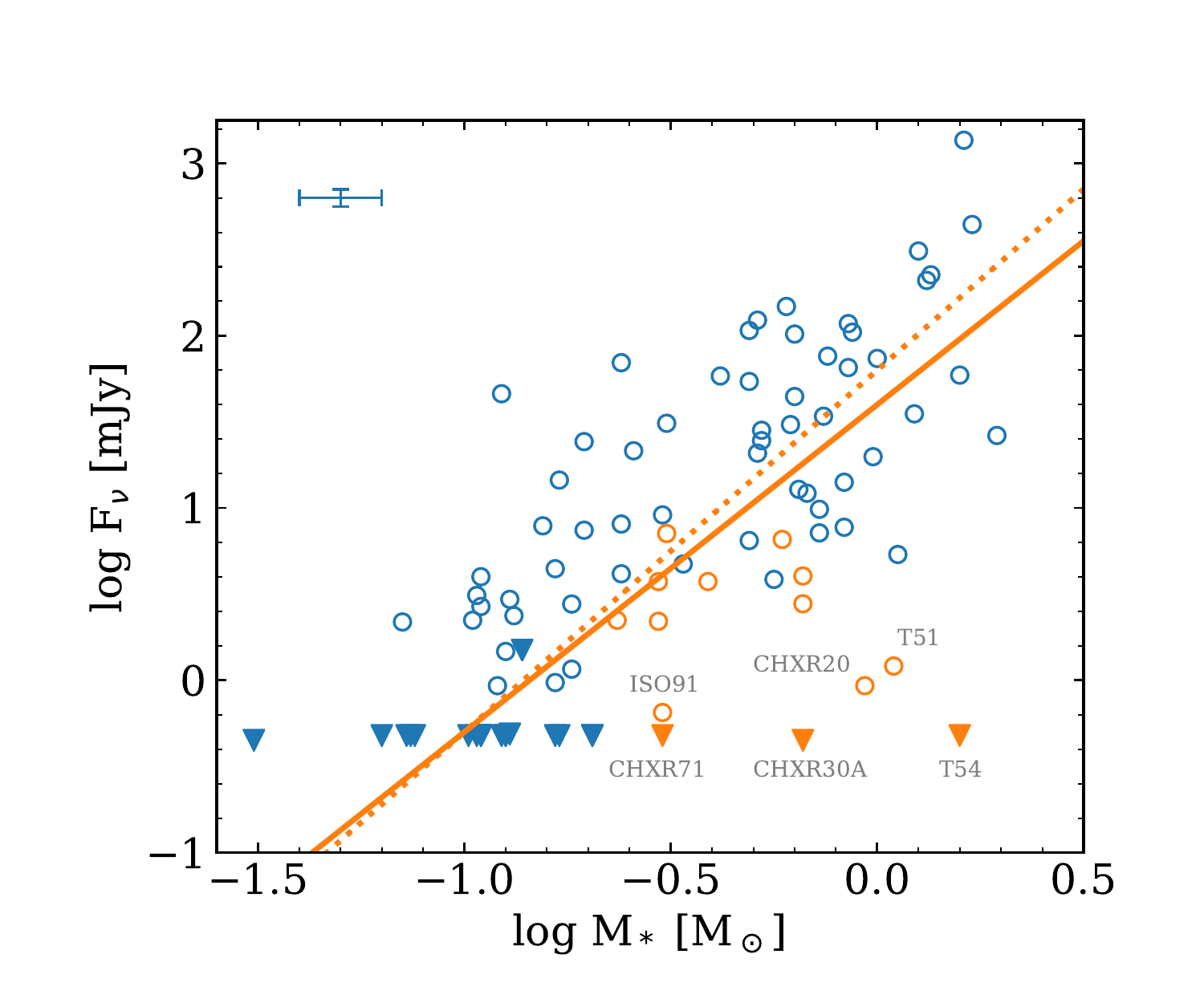}
    \includegraphics[width=0.5\textwidth, trim=0 0 0 0]{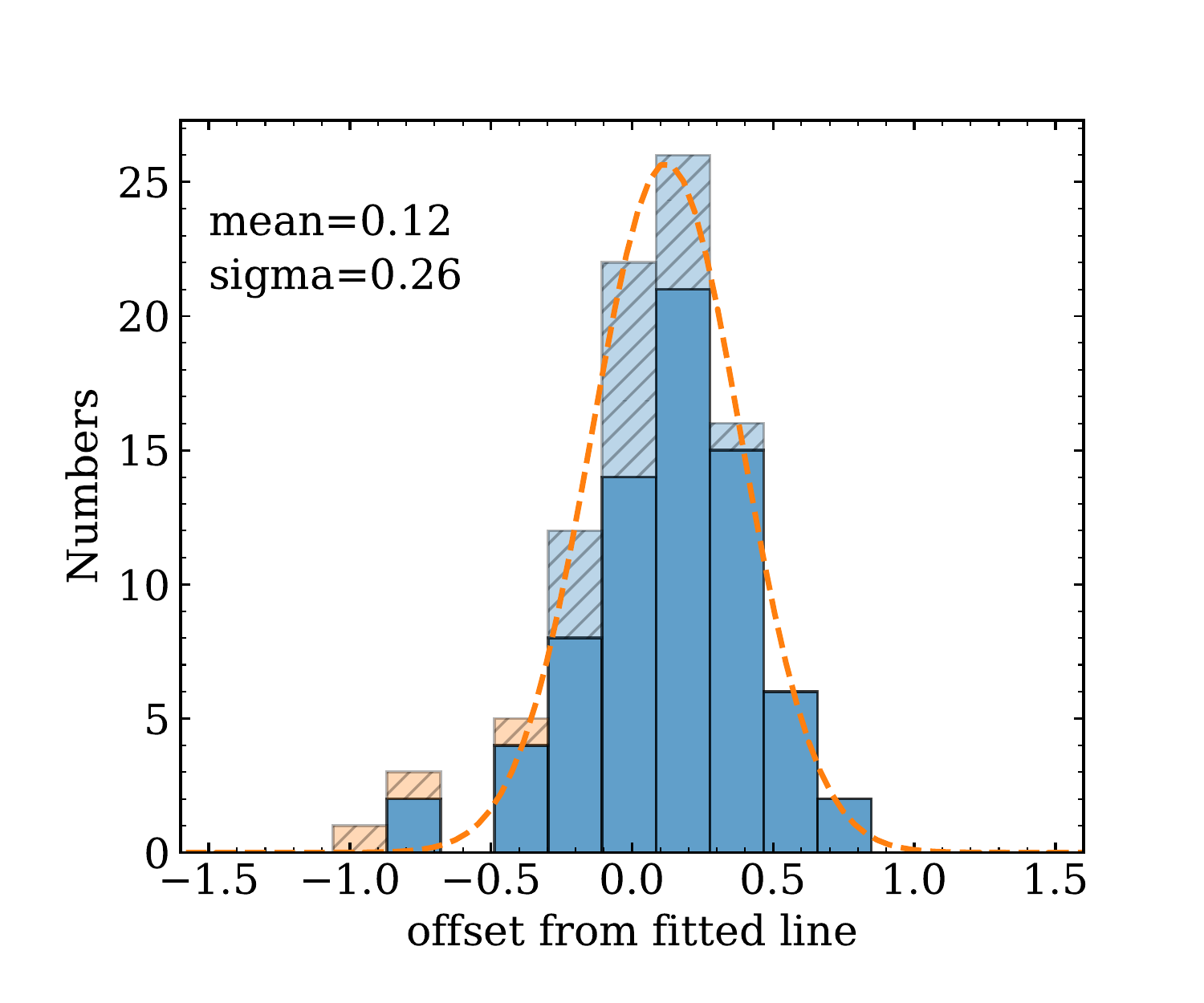}
    \caption{Left: Continuum flux densities as a function of stellar mass. The 14 sources re-observed in Cycle 3 are shown in orange, while the others are shown in blue with fluxes adopted from Cycle 2 observations \citep{pascucci2016}. Open circles for detections and downward triangles for non-detections. The best fit relation (orange full line) using updated fluxes and upper limits are highly consistent with relation (almost overlapped thus not shown here) from \citet{pascucci2016}. Revised relation without considerable outliers (see text for how they are selected) is shown in dotted line. The median error bars for log($M_*$) and log($F_{mm}$) are shown in the left upper corner and correspond to 0.1 dex and 0.05 dex; Right: Distribution of offsets from data points to the fitted line. Offsets in positive represent sources with higher continuum fluxes. Fitted parameters for the Gaussian distribution are labeled in text. The non-detections are shown in hatches to separate from the detections shown in full histogram, and the three Cycle 3 non-detections are shown in orange.  \label{fig:fmm_ms}}
\end{figure*}

Figure \ref{fig:fmm_ms} (left panel) presents the millimeter flux densities and 3$\sigma$ upper limits as a function of stellar mass in the log-log plane. This figure is similar to Figure 6 in \citet{pascucci2016}, but updated with the 14 new fluxes in our Cycle 3 sample.  
The best-fit correlation between disk flux and stellar mass is $\log(F_{mm}/{\rm mJy})=1.89 (\pm0.24)\times \log(M_\star/M_\odot)+1.57(\pm 0.14)$, assuming a linear scaling relation in the log-log plane and using a Bayesian linear regression method, following \citet{pascucci2016} \footnote{\citet{pascucci2016} used the \citet{kelly2007} $IDL$ routine, while here we use the python equivalent \textit{Linmix}, from https://github.com/jmeyers314/linmix.  We refer the reader to Appendix A in \citet{pascucci2016} for a description of how results from this method are consistent with results from a simpler linear regression.}. The fitted parameters are consistent with the best-fit values from \citet{pascucci2016}, with a slope of $1.9\pm0.2$ and an intercept of $1.6\pm0.1$ (see more details in \citet{pascucci2016} for the fitting method).

These relationships are consistent because the scaling relation fits assume that the data points are scattered in a Gaussian distribution.  With a higher detection rate, we confirm that this assumption is mostly correct, with some caveats.  Figure \ref{fig:fmm_ms} shows that the full range of disk mm flux is an order of magnitude at any given stellar mass. The offset between each data point and the fitted line shows a Gaussian fit with a mean value of 0.12 and 1$\sigma$ spread of 0.26 (see right panel of  Figure \ref{fig:fmm_ms}). The fitted positive mean value (or offset) is due to the inclusion of upper limits with low fluxes. This fit also reveals that four disks are outliers, with fluxes that are $>3\sigma$ fainter than the mean.

Since these four sources are anomalously faint, they may follow a different evolutionary path.  A better fitting approach may be to exclude them from the scaling relation.  When these four outliers, and a random sample of the same fraction (7\%) of the upper limits in the cool sample, are excluded from the fit, we obtain a $F_{mm}$ -- $M_\star$ scaling relation of $\log(F_{mm}/{\rm mJy})=2.07 (\pm0.21)\times \log(M_\star/M_\odot)+1.81(\pm 0.13)$. Therefore, excluding the outliers steepens the $F_{mm}$ -- $M_\star$ correlation by 1$\sigma$ and increases the average disk mass by $\sim 0.2$ dex in the \textit{Hot} sample, though consistent within 3$\sigma$ of the uncertainty reported in \citet{pascucci2016}.

The correlation between disk mass (mm flux) and stellar mass depends on the model of stellar evolution used to estimate stellar mass.  The stellar masses in this work are obtained from \citet{pascucci2016}, which were calculated by comparing stellar parameters (mostly from \citealt{manara2016} and \citealt{manara2017}) to a combination of the \citet{baraffe2015} models at low masses and the non-magnetic \citet{feiden2016} models\footnote{The non-magnetic models of \citet{feiden2016} were not available at the time of submission.} at higher masses.  Use of the \citet{feiden2016} magnetic models would increase the stellar masses\footnote{Both models produce masses that correlate well with stellar masses, as measured from disk rotation \citep{guilloteau2014,czekala2015,simon2017,yen2018} and young eclipsing binaries \citep[see compilation by][]{stassun2014}.  Use of the Feiden magnetic models would also increase the age of Chamaeleon I by $\sim 0.3$ dex.} by $\sim 0.2$ dex (between 0.1--0.9 $M_\odot$), thereby shifting the correlation to higher masses (i.e., the intercept in the best-fit correlation between disk and stellar mass would decrease by $\sim$0.4).
The updated distance has increased disk masses and stellar luminosities by 0.15 dex relative to \citet{pascucci2016}.  Since most stars in our sample are fully convective and evolving along the near-vertical Hayashi track, the mass is determined mostly by the spectral type (temperature) and is only minimally affected by the change in luminosity.

\begin{deluxetable*}{cccccc}
\tablecaption{Source Properties For Faint Disks \label{tab:faint-disk}}
\tablewidth{0pt}
\tablehead{
\colhead{Name} & \colhead{Multiplicity} &\colhead{Accretion} & \colhead{IR excess} & \colhead{mm detection} & \colhead{possible explanations}\\
\colhead{ } &\colhead{(arcsec) }  &\colhead{ } &\colhead{ } & \colhead{ } & \colhead{ }} 
\colnumbers
\startdata
CHXR71  &  0.56  &  No   &  weak &  No & tidal truncation / internal photoevaporation  \\
CHXR30A &  0.46  &  No  &  weak &  No  & tidal truncation / internal photoevaporation  \\
T54     &  0.24  &  No  &  weak &  No  & tidal truncation / internal photoevaporation  \\
CHXR20  & 28.46  &  Yes?    &  weak &   Yes & grain growth\\
T51     &  1.97  &  Yes?    &  normal &  Yes & tidal truncation\\
\enddata
\tablecomments{Accretion measurements in literature for CHXR20 and T51 show contradictory results, so marked with `?'. See more details in Appendix \ref{sec:faint-detail}.}

\end{deluxetable*}

\section{Faint disks}  \label{sec:outliers}
In the following section, we investigate the source properties of the faintest disks in our Cycle 3 sample. The four outliers identified in the previous section consist of two weak continuum detections (CHXR20 and T51) and two non-detections (CHXR30A and T54). Another two faint disks, ISO91 (a 3.8$\sigma$ detection) and CHXR71 (a non-detection), are not classified as anomalous outliers because of their lower stellar masses, but both have disk masses that are much lower than those of sources with similar stellar mass (see Figure \ref{fig:fmm_ms}) and are also included in this discussion. 

\begin{figure*}[t]
\centering
    \includegraphics[width=0.95\textwidth, trim=0 650 0 0]{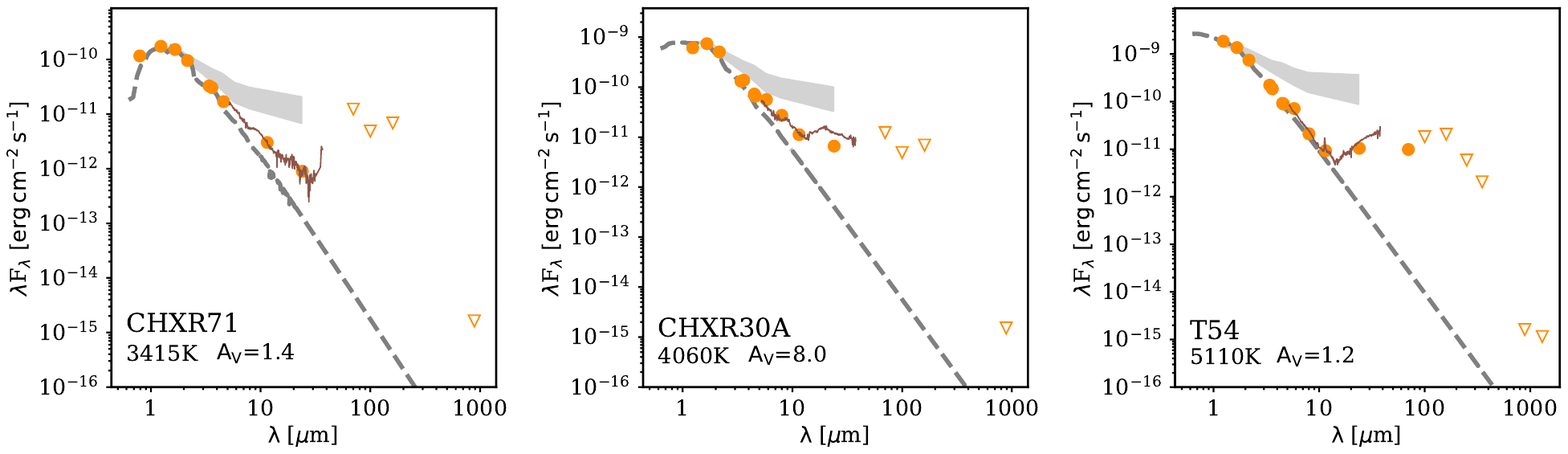}
    \caption{SEDs for the three sources in which 887 $\mu$m are not detected in our Cycle 3 observation. The $T_{eff}$ and $A_v$ are adopted from \citet{manara2017}. We show here the extinction-corrected optical and infrared photometry \citep{epchtein1999,skrutskie2006,luhman2007} , \textit{Spitzer} IRS spectra and ALMA upper limits. \textit{Herschel} upper limits are adopted from the average measurements or upper limits in Chamaeleon I samples \citep{olofsson2013}, except data for T54, which is adopted from \citet{matra2012}. We also include the 1.3 mm upper limits for T54 from \citet{hardy2015}. The light gray filled region shows the upper and lower quartiles of the median SEDs of L1641 ($\sim$1.5 Myr) CTTSs with similar spectral type to each source \citep{fang2013}, normalized at the \textit{H-}band flux of each source.  The dashed gray lines are the NextGen stellar model \citep{hauschildt1999,allard2000} at specified $T_{eff}$ which are normalized to \textit{J-}band flux.  \label{fig:sed_non}}
\end{figure*}

\begin{figure*}[!t]
\centering
    \includegraphics[width=0.95\textwidth, trim=0 650 0 0]{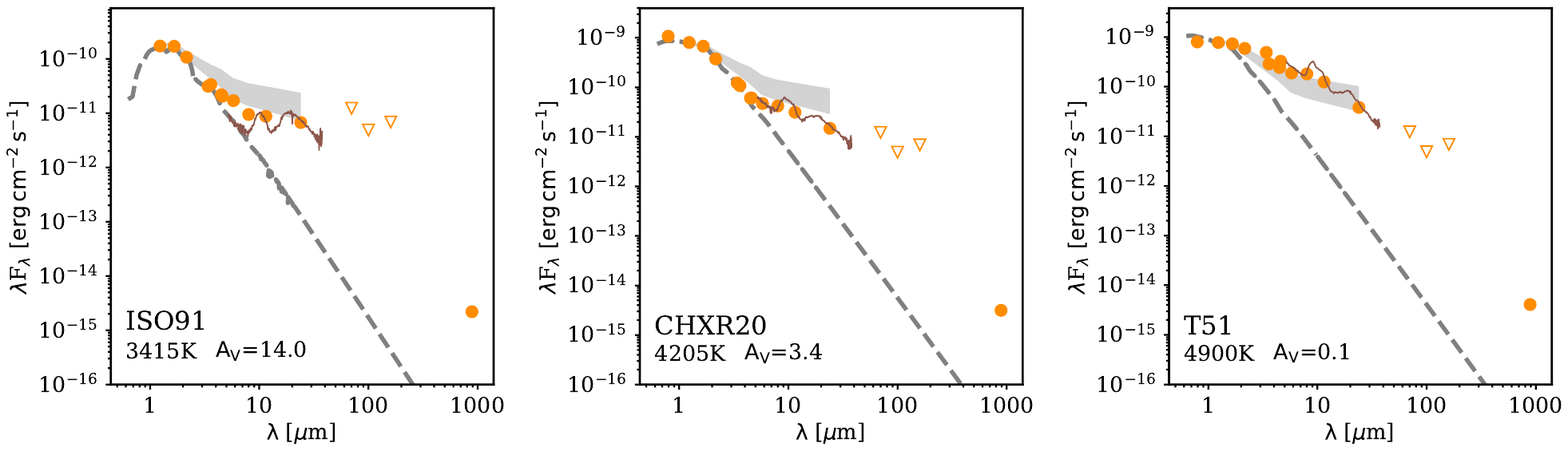}
    \caption{Same as Figure \ref{fig:sed_non}, but for the 3 sources in which 887 $\mu$m emission are only weakly detected. \label{fig:sed_weak}}
\end{figure*}

Properties of these six sources are summarized in Table \ref{tab:faint-disk} and are discussed in detail below (see also the Appendix \ref{sec:faint-detail} for a more detailed description of individual sources), including their SEDs (Figure \ref{fig:sed_non} and Figure \ref{fig:sed_weak}), accretion properties, and stellar multiplicity.  All six sources are classified as Class II objects based on their infrared spectral indices between 2 $\mu$m and 24 $\mu$m \citep{luhman2008}.

\subsection{Non-detections at mm wavelengths}
Figure \ref{fig:sed_non} presents the SEDs for the three sources (CHXR71, CHXR30A, and T54) that are not detected in our ALMA observations. Their SEDs show little or no excess continuum emission at wavelengths $\la$8 $\mu$m, indicating that their inner optically thick disks have mostly dissipated. The 10 $\mu$m silicate features in CHXR71 and T54 are weak or barely present, which suggests a deficit of small dust grains \citep{furlan2009} and/or a high degree of dust settling \citep{manoj2011}. The dust mass upper limits are only $\sim$0.2 $M_\earth$, comparable to the high mass end of dust masses in debris disks \citep{Moor2017}.

CHXR71, CHXR30A, and T54 have stellar companions located at projected separations of $0.\arcsec56$, $0.\arcsec46$, and $0.\arcsec24$ \citep{lafrenire2008,daemgen2013}, respectively, corresponding to 106 AU, 87 AU, and 45 AU
at the given distance of Chamaeleon I. In each system, the primary component dominates the UV and optical spectrum, from which the measured accretion rate is consistent with chromospheric activity \citep{manara2016,manara2017}. 

\subsection{Weak detections at mm wavelength}
The SEDs for the three weak detections (CHXR20, ISO91, and T51) are shown in Figure \ref{fig:sed_weak}. T51 has a near-IR excesse typical of many Class II disks, while CHXR20 shows no emission excess at wavelength $\la$6 $\mu$m. The 10 $\mu$m silicate emission features are detected in both systems. The low 13--31 $\mu$m spectral index in T51 is explained by \citet{furlan2009} and \citet{manoj2011} as an outward truncation by the $1.\arcsec9$ stellar companion \citep{lafrenire2008}. CHXR20, however, is only accompanied by a very wide (28.$\arcsec$46) companion \citep{kraus2007}, without any known close companion found from either high-resolution spectroscopy \citep{nguyen2012} or Adaptive Optics imaging \citep{lafrenire2008}. 
Optical spectra obtained with VLT/X-Shooter demonstrate that both components of T51 are accreting at a rate expected for their spectral type, while CHXR 20 is weakly accreting (\citealt{manara2016,manara2017}; see further details in the Appendix).

For ISO91, even though mm continuum emission is only detected at 4$\sigma$, the extreme extinction and the large-scale $^{12}$CO outflow (see $^{12}$CO channel maps in Appendix \ref{sec:faint-detail}) indicate that it may be at an earlier evolutionary stage, still surrounded by an envelope.  ISO91 is therefore excluded in the further analysis.

\section{Discussion} \label{sec:diss}
The sample of faint disks discussed above covers a wide range of SED morphologies, accretion rates, and binarity. This diversity may refer to different evolutionary stages and/or different mechanisms in shaping the disk. In the following sections, we will first compare source properties of these faint disks to the entire Chamaeleon I sample, and then attempt to place each disk on its own evolutionary track based on source properties.


\begin{figure}[!t]
    \includegraphics[width=0.5\textwidth, trim=0 0 0 0]{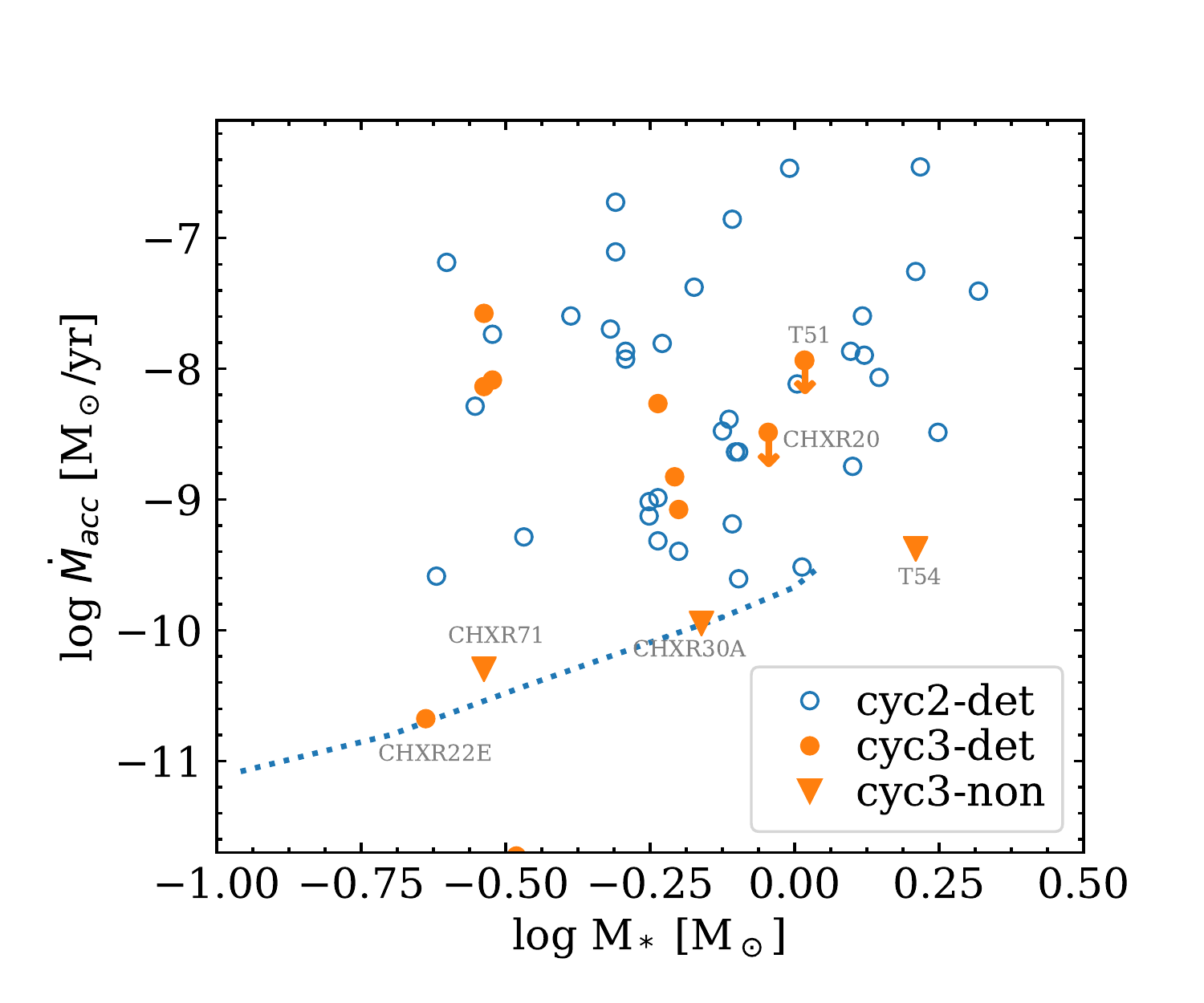}
    \caption{Mass accretion rate for the Cycle 2 and Cycle 3 detections are marked with blue open circles and orange dots, respectively. The typical upper limit on accretion rate versus stellar mass (dotted line) is calculated by \citet{manara2013} based on the intensity of line emission in non-accreting young stellar objects. All of the three millimeter non-detections are not accreting at detectable level.  
    \label{fig:accretion}}
\end{figure}

As a parallel effort to understand young stellar system evolution, 
\citet{manara2016,manara2017} measured mass accretion rates\footnote{The mass accretion rates used here are scaled to the new Gaia DR2 distance $d$ of 190 pc by multiplying by $\left(\frac{190 ~{\rm pc}}{160~{\rm pc}}\right)^3=1.67$, since the accretion rate is proportional to the accretion luminosity ($\propto d^2$) and stellar radius ($\propto d^1$).} for most sources in our ALMA Cycle 2 Chamaeleon I disk survey \citep{pascucci2016}, based on UV continuum excess, or line luminosity when signal-to-noise ratio at short wavelength is poor. Accretion rates 
for the three non-detections (CHXR71, CHXR30A, and T54) in the \textit{Hot} sample are all below or close to the expected emission from chromosphere (see Figure \ref{fig:accretion})\footnote{2MASS J10533978-7712338 is located far below the plot boundary and is likely an edge-on disk \citep{luhman2007}, thus its stellar properties and measured accretion rate are highly uncertain and not trustworthy, and is excluded from the plot.}, indicating that any on-going accretion is weak.  The uncertainties in accretion rate are typically $\sim$0.3 dex and are more uncertain near the chromospheric boundary\footnote{\citet{herczeg2008}, \citet{alcala2014}, and \citet{manara2016} provide detailed discussions on the uncertainties in these measurements, while \citet{costigan2014} and \citet{venuti2015} quantify variability in accretion rate measurements.}. The other two faint sources, CHXR20 and T51 (marked with a downward arrow in Figure \ref{fig:accretion}), have contradictory accretion rate measurements in the literature but are established by X-Shooter spectra to be accreting \citep[e.g.][]{manara2016,manara2017}. Meanwhile, the vast majority of continuum detections have accretion rates much higher than the typical chromospheric emission. CHXR22E, however, stands out as an exception with detected millimeter emission and no detected accretion. CHXR22E is classified as a transitional disk candidate in \citet{kim2009} without a near-infrared excess (also see Figure \ref{fig:sed_bright}).  CHXR22E also has no reported close stellar companion \citep{lafrenire2008}.  Since accretion is variable and our accretion rates are from single epochs, the rates do not necessarily correspond to the time-averaged accretion onto these objects.

\begin{figure}
    \includegraphics[width=0.5\textwidth, trim=0 0 0 0]{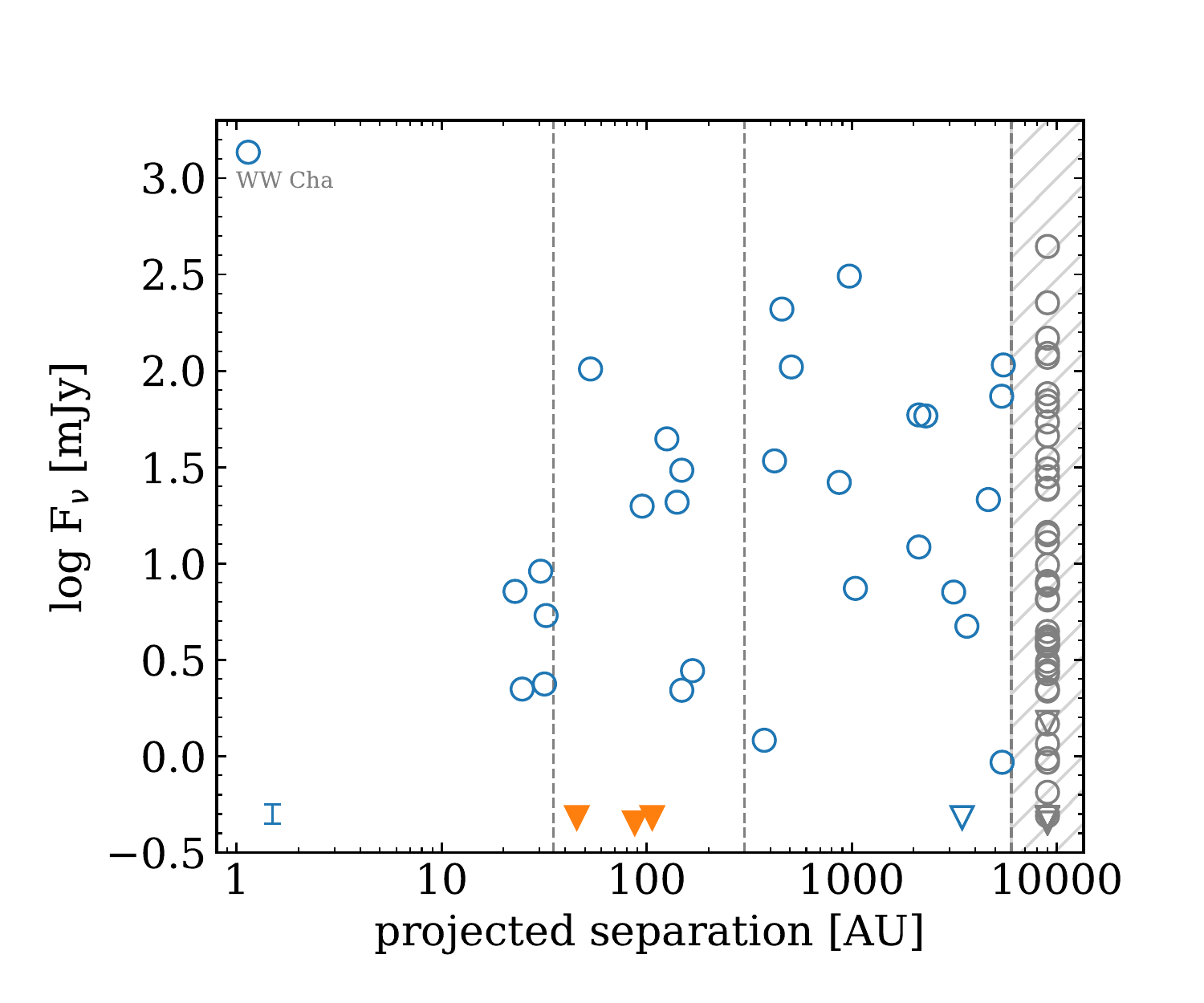}
    \caption{millimeter flux densities as a function of projected binary separation. Single stars are shown in the right hatched region for reference. The three \textit{Hot} sample non-detections are highlighted in orange downward triangles. The two vertical gray dashed lines represent 30 AU and 300 AU, as proposed in \citet{harris2012} to group the Taurus sample. The median error bar for log($F_{mm}$) is 0.05 dex and shown in the left lower corner. \label{fig:binary}}
\end{figure}

The three non-detections are close binaries with projected separations of 40--100 AU. The weakly detected source, T51, also has a stellar companion separated by at least $\sim$360 AU. Based on unresolved disk diagnostics, disk lifetimes in close binaries (with separation $\la$ 40 AU) are shorter than lifetimes of disks around single stars \citep{bouwman2006,cieza2009,kraus2012,daemgen2016}. From component-resolved millimeter observations in a large sample of disks in Taurus, flux densities decrease significantly as a function of decreasing pair separation \citep{harris2012}. 
More specifically, \citet{harris2012} found substantially lower flux densities at projected separations of 30 AU and 300 AU.
As seen from Figure \ref{fig:binary}, this trend is not prominent in Chamaeleon I, though binary systems with small separations ($\la$ 30 AU, if excluding the spectroscopic binary -- WW Cha) have lower average millimeter flux. The spectroscopic binary, WW Cha (marked in Figure \ref{fig:binary}), is the brightest continuum source in our sample \citep{pascucci2016}, consistent with the trend found in \citet{harris2012} that circumbinary disks around spectroscopic binaries tend to have higher fluxes.
We note that systems with projected separations $\la$ 100 AU should remain unresolved at our spatial resolution. If both individual disks exist, millimeter fluxes for small separation systems should be lower than the value reported here. 

In the following subsections, we discuss physical mechanisms that could explain the evolutionary stage of these mm-faint sources. 
We start with possible explanations for individual disk evolutions, since our binaries are largely unresolved and most likely to retain their own circumstellar disks at a separation of $\sim$100 AU. We then discuss disk evolution in binary systems and the possibility of a disk around the fainter secondary star.

\subsection{Individual disk evolution}   
As disk mass and accretion rate decrease over time, photoevaporation becomes the leading role in draining the inner disk and eventually creates inner cavities \citep{alexander2014}. Once the inside-out dissipation begins, accretion ceases quickly as material from the outer disk can not pass through the photoevaporation front. Observationally, most WTTSs show little or no near-IR excess at wavelength $\la$6 $\mu$m \citep{padgett2006,mccabe2006,manoj2011}. The three disks in this paper that are not detected at 887 $\mu$m are also not accreting at detectable level and have SEDs that indicate some clearing of the inner disk, both consistent with internal photoevaporation. \citet{kim2009} modeled an inner dust disk cavity of 37 AU for T54, which is broadly consistent with the predicted size of the inner disk hole from photoevaporation models driving by EUV+X-ray radiation at its given upper limit of accretion rate \citep[see more details in Figure 6 of][]{ercolano2017}. Comparisons with photoevaporation models for CHXR71 and CHXR30A are limited by the lack of previous studies.

Starting from a primordial flared disk, larger particles formed by collisional coagulation feel less thermal pressure support and settle towards the disk mid-plane \citep{weidenschilling1977}. Fast grain growth is expected in the denser regions through which small particles are gradually depleted, thereby leading to a strong decrease of flux densities from near to mid-infrared wavelengths \citep{dullemond2005,dalessio2006}. This overall decrease in the level of excess emission across the disk SED resembles the so-called ``evolved" disk \citep{lada2006,hernandez2007,currie2009,luhman2010}. A wide range of accretion rate and disk mass are reported for such disks \citep{cieza2012}. Instead of a sharp rise in the mid-infrared wavelength region ($>$10 $\mu$m), as is typical for  transition disks, CHXR20 and the three non-detection sources show weaker mid-infrared excesses. Fast grain growth is capable of explaining the observed SED morphologies in these four disks, indicating disk evolution by grain growth instead of direct disk dissipation.

\subsection{Disk evolution in binary systems}
A stellar companion orbiting the primary star, separated by a typical disk radius, can dramatically alter the disk structure and its evolution. Strong tidal interactions truncate the disk radii and reduce disk masses \citep{jensen1994,jensen1996,harris2012,akeson2014}. Theoretical models predict the fate for each of the three disks in a binary system (i.e., circumprimary, circumsecondary, and circumbinary) in a range of stellar and disk configurations (mass ratio $q$, semi-major axis $a$,  orbital eccentricity $e$ and disk viscosity $\alpha$) \citep{papaloizou1977,artymowicz1994,miranda2015,lubow2015}. The classical calculation for a co-planar binary in \citet{artymowicz1994} predicts an outer truncation for circumstellar disks at $\sim$0.3$a$ and an inner truncation for circumbinary disks of $\sim$2--3$a$, though disk truncation for misaligned systems are not fully explored \citep[e.g.,][]{miranda2015}.

According to disk truncation models, the $\sim$2$\arcsec$ companion in T51 would have truncated the outer circumstellar disk at a radius of $\sim$100--150 AU. Given the large beam size ($\sim$100 AU in radius) in our observations, the disk size is not well constrained. 

Stellar companions for the three non-detections, T54, CHXR30A, and CHXR71, are separated in projected distance by 45, 87, and 106 AU, respectively.
Individual disks must have undergone tidal stirring in these systems. Given the binary separations, circumbinary disks are more plausible to be present for T54 than for CHXR71 and CHXR30A. \citet{espaillat2017} reproduced the IR SED of T54 with a model consisting of a circumprimary disk ($\sim$10 AU) and a circumbinary disk (60--100 AU). Tidal truncation in the outer disk and photoevaporation in the inner edge could cooperate together to explain non-detections in both millimeter emission and accretion for the three sources.

Close companions ($\la$40--100 AU) can shorten disk lifetimes \citep{cieza2009,kraus2012} and reduce disk masses \citep{harris2012,cox2017}, which should have direct consequences on planets formed within. Despite such a hostile environment, many planets have been detected in circumstellar orbits of binary systems in a wide range of separations \citep[e.g.,][]{eggenberger2007}, in addition to extreme cases like multiple rocky planets in binaries (e.g., Kepler-444, \citealt{dupuy2016}). 
Though the overall binary fraction of \textit{Kepler} exoplanet host stars is estimated to be comparable to the general field population ($\sim$40\%) \citep{horch2014}, planet formation is strongly suppressed around binary hosts with separation from $\sim$20 AU up to 100 AU \citep{eggenberger2007,bonavita2007,wang2014a,kraus2016}. 
This difference is expected to be a direct consequence of the faster disk dispersal in binary systems, though grain growth in binary systems with tens of AU separations is indistinguishable from single stars \citep{pascucci2008}. The above statistics only applies to circumstellar planets, leaving circumbinary planets untouched.

\subsection{Disks around the secondary star} 
In unequal-mass binary systems, the disk could survive longer around the secondary star than around the primary star, as has been found for several systems with mid-IR imaging \citep{mccabe2006}.  Disks around cool stars and sun-like stars are quite different in physical and chemical structures \citep[e.g.,][]{pascucci2009,szucs2010},
though similar dust properties are able to explain the observed median IR SEDs across a range of stellar mass \citep{mulders2012}.  Very low mass stars and brown dwarfs have flatter disks and overall less excess at IR bands \citep[e.g.,][]{apai2005,liu2015}, with SEDs similar to Figure \ref{fig:sed_non}. If the disk is around the fainter secondary star, the weaker radiation field should result in less emission at all IR bands, compared to the case of disk around the primary star. In unresolved observations, the photospheric emission from the primary star would dominate at all short wavelengths, until somewhere in the mid-IR where the dust is brighter. Accretion from the secondary star is therefore hard to be detectable in the optical spectra.

\begin{figure}[!t]
    \includegraphics[width=0.5\textwidth, trim=0 0 0 0]{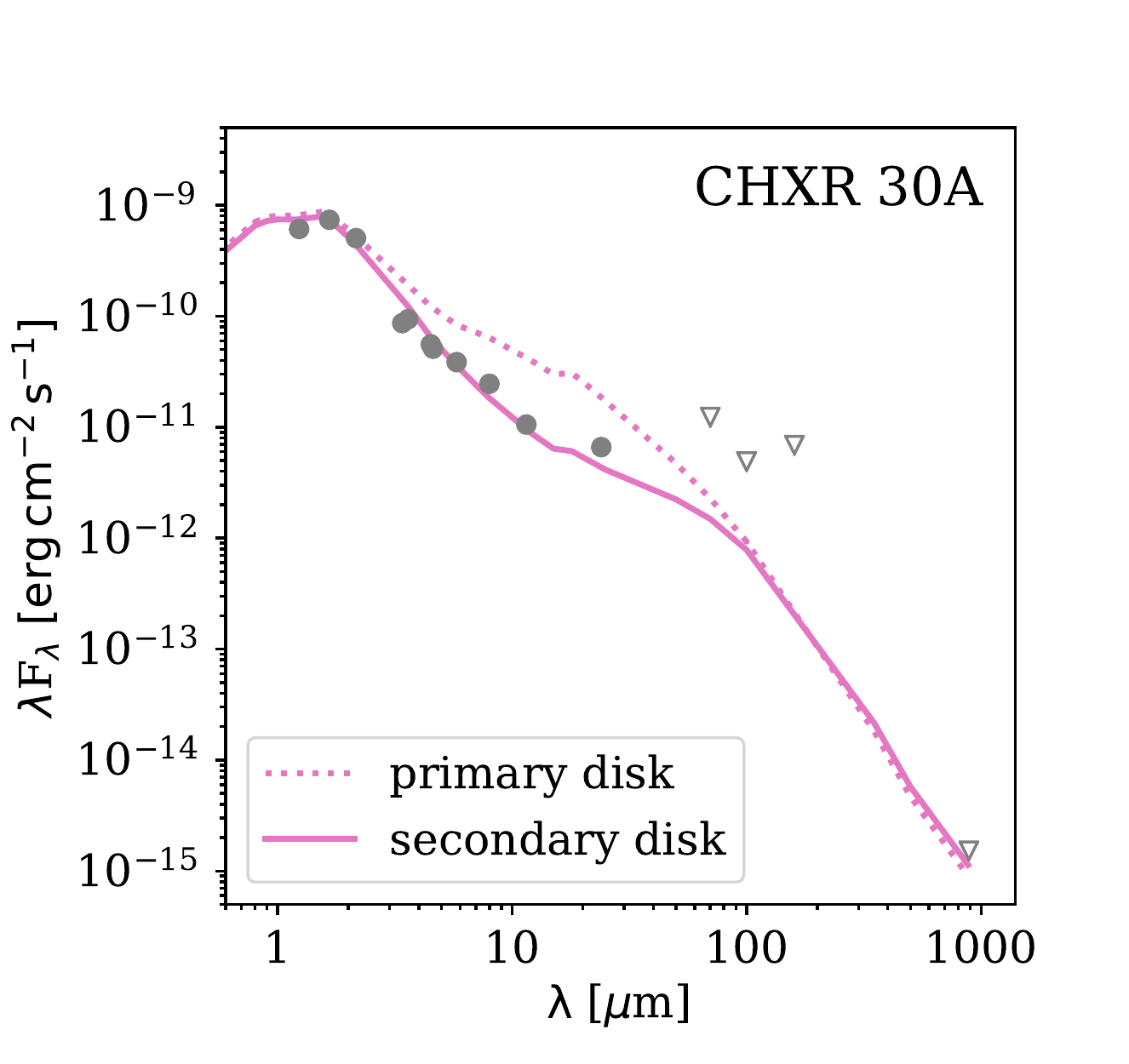}
    \caption{Modeled SED for CHXR30A. Solid line represents scenario of disk around the faint secondary, while dotted line for disk around the primary. \label{fig:radmc}}
\end{figure}

If the disk is around the secondary star, the disk may be weak or undetected in the sub-mm because (a) disks around lower-mass objects are expected to be less massive (Figure \ref{fig:fmm_ms}), and (b) the disk will also be colder than assessed because of the weaker radiation field of the lower-mass star.
To test the possibility of a disk around the secondary star in a binary system, we use the radiative transfer code RADMC-3D \citep{dullemond2012} to model the SED of the binary system CHXR30A.
Stellar properties for the primary star are adopted from \citet{manara2017}. The $K_s$-band contrast of 0.85 from \citep{lafrenire2008} leads to an estimated stellar mass of 0.28 $M_\odot$ \footnote{The mass and luminosity are actually upper limits, since these calculations assume that the $K$-band emission from the secondary star does not include any disk emission.}.
Based on the projected separation, the two stellar components are placed at 80 AU apart.  The secondary disk is therefore truncated at $\sim$30 AU\footnote{The truncation radius could be lower limit, if the real binary separation is wider than the projected separation.} , following predictions from the evolution of disks in binaries \citep{artymowicz1994}. The ratio of disk mass \footnote{In the disk model, we include two populations of amorphous dust grains with 25\% carbon and 75\% silicate, resulting in a dust opacity a few times higher than what used in disk dust mass calculation \citep{pascucci2016}.} to stellar mass is adjusted to satisfy the millimeter flux upper limit.

The modeled SED reproduces well the observed SED structure, as shown in Figure \ref{fig:radmc}.  In this scenario, the dust continuum emission at 887 $\mu$m is still fainter, by a factor of few than expected from the stellar mass-disk mass correlations, due to a truncated outer disk.  If we keep most of the disk parameters fixed, but place the disk around the primary star and reduce the disk mass to satisfy the mm flux upper limit, 
the emission at IR wavelength would be brighter by an order of magnitude.

While identifying the best-fit model for these systems is beyond the scope of this paper, this experiment verifies the possibility that a circumsecondary disk may explain the observed stellar SEDs of our mm-faint binary systems.  Binary systems, where the disk is around the secondary star, may confuse models of disk evolution because these systems are rare enough to not dominate the population statistics, but are common enough to be identified as outliers that could link different evolutionary states.  For example, the star CZ Tau AB is a $0\farcs3$ M3+$\sim$M6 binary system with an SED with no near-IR excess, a sharp rise from 5--20 $\mu$m, then a rapid falloff at longer wavelengths \citep{furlan2006} and a non-detection in the sub-mm \citep{andrews2013}. The optical spectrum shows no accretion from the primary \citep{herczeg2014}, but the secondary is faint enough to not contribute significantly to optical spectra.  The K- and L-band imaging of \citet{mccabe2006} shows that the disk is likely around the secondary, a simple scenario that \citep[along with disk truncation in][]{furlan2006} explains all of the existing observations without requiring any speculation about the CZ Tau disk being in a unique evolutionary stage of disk dispersal.

\citet{rosotti2018} predict that disks around the secondary in wide binary systems ($\ga$30 AU) could be longer lived than disks around the primary, possibly due to the lower X-ray photoevoparation rate around lower mass stars.  ALMA observations have also discovered binary systems with mm-brighter circumsecondary disk than circumprimary disk \citep{akeson2014}. Given the binary separation, the scenario of disk around the fainter sencondary is more likely to explain the observations for CHXR30A and CHXR71 than for the closer T54 system.

Current ALMA disk surveys of disk mass in near-complete or unbiased samples in nearby regions have spatial resolutions of $\sim$60--100 AU, leaving disks in multiple systems unresolved. High-resolution ALMA observations towards large binary samples in different star-forming regions are needed to draw a comprehensive picture of disk demographics in multiple star/disk systems. Any correlation of circumstellar or circumbinary disk to stellar mass ratio, binary separation or even the individual disk mass/size, if they exist, are worthy of an in-depth investigation for a better understanding of planetary architectures.

\section{Summary} \label{sec:sum}
In this paper, we presented 887 $\mu$m ALMA Cycle 3 observations for 14 Class II disks in the nearby $\sim$2 Myr-old Chamaeleon I star-forming region. These targets were selected based on spectral type (earlier than or equal to M3), the presence of a disk in mid-IR observations, and having a faint detection or non-detection in our ALMA Cycle 2 program \citep{pascucci2016}.  We have identified mm-faint outliers in the high-sensitivity data and explored source properties from IR SED, accretion and stellar multiplicity. Our main findings are as follows:

\begin{enumerate}
\item By improving the sensitivity by a factor of 5, we identify six more sources with dust continuum emission and increase the detection rate to 94$\%$ (51/54) in the Chamaeleon I \textit{Hot} sample. 

\item We find that the stellar-disk mass relation reported in \citet{pascucci2016} is robust to the inclusion of updated millimeter fluxes and upper limits for the faint sources. We identify mm-faint outliers which are located far from the main sample in the $F_{mm}$ -- $M_{*}$ plot. Excluding these outliers yields a $F_{mm}$ -- $M_{*}$ relation that is  1$\sigma$ steeper and an average disk mass that is $\sim 0.2$ dex higher.

\item The three millimeter non-detections (CHXR71, CHXR30A, and T54) show little or no NIR excess and weaker MIR excess than typical Class II objects. These stars have mostly stopped accreting and are all accompanied by a stellar companion at projected separation $\la$100 AU. We suggest that the combined effects of tidal interactions and internal photoevaporation hastens the overall disk evolution in these systems. The other three weak detections (ISO91, CHXR20, and T51) have a wide variety of source properties. CHXR20 might have undergone fast grain growth which depletes small particles especially $\mu$m-sized particles in the near to mid-IR wavelength and even grow to cm-size.  The faint millimeter flux for T51 could also be explained by a disk around the $\sim$2$\arcsec$ companion.

\item We also propose a scenario in which only a disk around the secondary star is left in a binary system to explain our observations.  For CHXR30A, the model SED that includes only a circumsecondary disk matches the observations quite well.  Such configurations may be common for binaries with separations $\sim$30--100 AU.

\end{enumerate}

Binarity information is critical when evaluating rare objects, such as weak-lined T Tauri stars with disks \citep[e.g.,][]{wahhaj2010}, both because of the exciting role that binarity may play in disk evolution \citep[e.g.,][]{kraus2012}, but also because disk evolution may be confused by a circumsecondary disk. Stellar multiplicity contributes significantly in explaining the wide range of disk mm fluxes at any given stellar mass bin in a single cluster, but it is definitely not enough for the whole picture. Given the complexity in star+disk system, a better understanding of disk fundamental properties (e.g., disk size, inclination, dust grain properties) is needed as the next first step in characterizing disk evolution.

\acknowledgments
We would like to thank the anonymous referee for providing constructive comments that improved the quality of this manuscript.  F.L. thanks Min Fang and Yao Liu for their help in RADMC modeling. F.L. and G.J.H are supported by general grants 11473005 and 11773002 awarded by the National Science Foundation of China. I.P. and N.H. acknowledges support from an NSF Astronomy \& Astrophysics Research Grant ( ID: 1515392). 

This paper makes use of the following ALMA data: ADS/JAO.ALMA\#2015.1.00333.S. ALMA is a partnership of ESO (representing its member states), NSF (USA) and NINS (Japan), together with NRC (Canada) and NSC and ASIAA (Taiwan), in cooperation with the Republic of Chile. The Joint ALMA Observatory is operated by ESO, AUI/NRAO and NAOJ.


\begin{thebibliography}{}
\expandafter\ifx\csname natexlab\endcsname\relax\def\natexlab#1{#1}\fi
\providecommand{\url}[1]{\href{#1}{#1}}

\bibitem[{{Akeson} \& {Jensen}(2014)}]{akeson2014}
{Akeson}, R.~L., \& {Jensen}, E.~L.~N. 2014, \apj, 784, 62

\bibitem[{{Alcal{\'a}} {et~al.}(2014){Alcal{\'a}}, {Natta}, {Manara}, {Spezzi},
  {Stelzer}, {Frasca}, {Biazzo}, {Covino}, {Randich}, {Rigliaco}, {Testi},
  {Comer{\'o}n}, {Cupani}, \& {D'Elia}}]{alcala2014}
{Alcal{\'a}}, J.~M., {Natta}, A., {Manara}, C.~F., {et~al.} 2014, \aap, 561, A2

\bibitem[{{Alexander} {et~al.}(2014){Alexander}, {Pascucci}, {Andrews},
  {Armitage}, \& {Cieza}}]{alexander2014}
{Alexander}, R., {Pascucci}, I., {Andrews}, S., {Armitage}, P., \& {Cieza}, L.
  2014, Protostars and Planets VI, 475

\bibitem[{{Allard} {et~al.}(2000){Allard}, {Hauschildt}, \&
  {Schweitzer}}]{allard2000}
{Allard}, F., {Hauschildt}, P.~H., \& {Schweitzer}, A. 2000, \apj, 539, 366

\bibitem[{{Andrews} {et~al.}(2013){Andrews}, {Rosenfeld}, {Kraus}, \&
  {Wilner}}]{andrews2013}
{Andrews}, S.~M., {Rosenfeld}, K.~A., {Kraus}, A.~L., \& {Wilner}, D.~J. 2013,
  \apj, 771, 129

\bibitem[{{Ansdell} {et~al.}(2017){Ansdell}, {Williams}, {Manara}, {Miotello},
  {Facchini}, {van der Marel}, {Testi}, \& {van Dishoeck}}]{ansdell2017}
{Ansdell}, M., {Williams}, J.~P., {Manara}, C.~F., {et~al.} 2017, \aj, 153, 240

\bibitem[{{Ansdell} {et~al.}(2016){Ansdell}, {Williams}, {van der Marel},
  {Carpenter}, {Guidi}, {Hogerheijde}, {Mathews}, {Manara}, {Miotello},
  {Natta}, {Oliveira}, {Tazzari}, {Testi}, {van Dishoeck}, \& {van
  Terwisga}}]{ansdell2016}
{Ansdell}, M., {Williams}, J.~P., {van der Marel}, N., {et~al.} 2016, \apj,
  828, 46

\bibitem[{{Apai} {et~al.}(2005){Apai}, {Pascucci}, {Bouwman}, {Natta},
  {Henning}, \& {Dullemond}}]{apai2005}
{Apai}, D., {Pascucci}, I., {Bouwman}, J., {et~al.} 2005, Science, 310, 834

\bibitem[{{Artymowicz} \& {Lubow}(1994)}]{artymowicz1994}
{Artymowicz}, P., \& {Lubow}, S.~H. 1994, \apj, 421, 651

\bibitem[{{Baraffe} {et~al.}(2015){Baraffe}, {Homeier}, {Allard}, \&
  {Chabrier}}]{baraffe2015}
{Baraffe}, I., {Homeier}, D., {Allard}, F., \& {Chabrier}, G. 2015, \aap, 577,
  A42

\bibitem[{{Barenfeld} {et~al.}(2016){Barenfeld}, {Carpenter}, {Ricci}, \&
  {Isella}}]{barenfeld2016}
{Barenfeld}, S.~A., {Carpenter}, J.~M., {Ricci}, L., \& {Isella}, A. 2016,
  \apj, 827, 142

\bibitem[{{Bonavita} \& {Desidera}(2007)}]{bonavita2007}
{Bonavita}, M., \& {Desidera}, S. 2007, \aap, 468, 721

\bibitem[{{Bouwman} {et~al.}(2006){Bouwman}, {Lawson}, {Dominik}, {Feigelson},
  {Henning}, {Tielens}, \& {Waters}}]{bouwman2006}
{Bouwman}, J., {Lawson}, W.~A., {Dominik}, C., {et~al.} 2006, \apjl, 653, L57

\bibitem[{{Calvet} {et~al.}(2005){Calvet}, {D'Alessio}, {Watson},
  {Franco-Hern{\'a}ndez}, {Furlan}, {Green}, {Sutter}, {Forrest}, {Hartmann},
  {Uchida}, {Keller}, {Sargent}, {Najita}, {Herter}, {Barry}, \&
  {Hall}}]{calvet2005}
{Calvet}, N., {D'Alessio}, P., {Watson}, D.~M., {et~al.} 2005, \apjl, 630, L185

\bibitem[{{Carilli} \& {Holdaway}(1999)}]{carilli1999}
{Carilli}, C.~L., \& {Holdaway}, M.~A. 1999, Radio Science, 34, 817

\bibitem[{{Cieza} {et~al.}(2012){Cieza}, {Schreiber}, {Romero}, {Williams},
  {Rebassa-Mansergas}, \& {Mer{\'{\i}}n}}]{cieza2012}
{Cieza}, L.~A., {Schreiber}, M.~R., {Romero}, G.~A., {et~al.} 2012, \apj, 750,
  157

\bibitem[{{Cieza} {et~al.}(2009){Cieza}, {Padgett}, {Allen}, {McCabe},
  {Brooke}, {Carey}, {Chapman}, {Fukagawa}, {Huard}, {Noriga-Crespo},
  {Peterson}, \& {Rebull}}]{cieza2009}
{Cieza}, L.~A., {Padgett}, D.~L., {Allen}, L.~E., {et~al.} 2009, \apjl, 696,
  L84

\bibitem[{{Clarke} {et~al.}(2001){Clarke}, {Gendrin}, \&
  {Sotomayor}}]{clarke2001}
{Clarke}, C.~J., {Gendrin}, A., \& {Sotomayor}, M. 2001, \mnras, 328, 485

\bibitem[{{Correia} {et~al.}(2006){Correia}, {Zinnecker}, {Ratzka}, \&
  {Sterzik}}]{correia2006}
{Correia}, S., {Zinnecker}, H., {Ratzka}, T., \& {Sterzik}, M.~F. 2006, \aap,
  459, 909

\bibitem[{{Costigan} {et~al.}(2014){Costigan}, {Vink}, {Scholz}, {Ray}, \&
  {Testi}}]{costigan2014}
{Costigan}, G., {Vink}, J.~S., {Scholz}, A., {Ray}, T., \& {Testi}, L. 2014,
  \mnras, 440, 3444

\bibitem[{{Covino} {et~al.}(1997){Covino}, {Alcala}, {Allain}, {Bouvier},
  {Terranegra}, \& {Krautter}}]{covino1997}
{Covino}, E., {Alcala}, J.~M., {Allain}, S., {et~al.} 1997, \aap, 328, 187

\bibitem[{{Cox} {et~al.}(2017){Cox}, {Harris}, {Looney}, {Chiang}, {Chandler},
  {Kratter}, {Li}, {Perez}, \& {Tobin}}]{cox2017}
{Cox}, E.~G., {Harris}, R.~J., {Looney}, L.~W., {et~al.} 2017, \apj, 851, 83

\bibitem[{{Currie} {et~al.}(2009){Currie}, {Lada}, {Plavchan}, {Robitaille},
  {Irwin}, \& {Kenyon}}]{currie2009}
{Currie}, T., {Lada}, C.~J., {Plavchan}, P., {et~al.} 2009, \apj, 698, 1

\bibitem[{{Czekala} {et~al.}(2015){Czekala}, {Andrews}, {Jensen}, {Stassun},
  {Torres}, \& {Wilner}}]{czekala2015}
{Czekala}, I., {Andrews}, S.~M., {Jensen}, E.~L.~N., {et~al.} 2015, \apj, 806,
  154

\bibitem[{{Daemgen} {et~al.}(2016){Daemgen}, {Elliot Meyer}, {Jayawardhana}, \&
  {Petr-Gotzens}}]{daemgen2016}
{Daemgen}, S., {Elliot Meyer}, R., {Jayawardhana}, R., \& {Petr-Gotzens}, M.~G.
  2016, \aap, 586, A12

\bibitem[{{Daemgen} {et~al.}(2013){Daemgen}, {Petr-Gotzens}, {Correia},
  {Teixeira}, {Brandner}, {Kley}, \& {Zinnecker}}]{daemgen2013}
{Daemgen}, S., {Petr-Gotzens}, M.~G., {Correia}, S., {et~al.} 2013, \aap, 554,
  A43

\bibitem[{{D'Alessio} {et~al.}(2006){D'Alessio}, {Calvet}, {Hartmann},
  {Franco-Hern{\'a}ndez}, \& {Serv{\'{\i}}n}}]{dalessio2006}
{D'Alessio}, P., {Calvet}, N., {Hartmann}, L., {Franco-Hern{\'a}ndez}, R., \&
  {Serv{\'{\i}}n}, H. 2006, \apj, 638, 314

\bibitem[{{Duch{\^e}ne} \& {Kraus}(2013)}]{duchene2013}
{Duch{\^e}ne}, G., \& {Kraus}, A. 2013, \araa, 51, 269

\bibitem[{{Dullemond} \& {Dominik}(2005)}]{dullemond2005}
{Dullemond}, C.~P., \& {Dominik}, C. 2005, \aap, 434, 971

\bibitem[{{Dullemond} {et~al.}(2012){Dullemond}, {Juhasz}, {Pohl}, {Sereshti},
  {Shetty}, {Peters}, {Commercon}, \& {Flock}}]{dullemond2012}
{Dullemond}, C.~P., {Juhasz}, A., {Pohl}, A., {et~al.} 2012, {RADMC-3D: A
  multi-purpose radiative transfer tool}, Astrophysics Source Code Library, , ,
  ascl:1202.015

\bibitem[{{Dunham} {et~al.}(2015){Dunham}, {Allen}, {Evans},
  {Broekhoven-Fiene}, {Cieza}, {Di Francesco}, {Gutermuth}, {Harvey},
  {Hatchell}, {Heiderman}, {Huard}, {Johnstone}, {Kirk}, {Matthews}, {Miller},
  {Peterson}, \& {Young}}]{dunham2015}
{Dunham}, M.~M., {Allen}, L.~E., {Evans}, Neal~J., I., {et~al.} 2015, The
  Astrophysical Journal Supplement Series, 220, doi:10.1088/0067-0049/220/1/11

\bibitem[{{Dupuy} {et~al.}(2016){Dupuy}, {Kratter}, {Kraus}, {Isaacson},
  {Mann}, {Ireland}, {Howard}, \& {Huber}}]{dupuy2016}
{Dupuy}, T.~J., {Kratter}, K.~M., {Kraus}, A.~L., {et~al.} 2016, \apj, 817, 80

\bibitem[{{Durisen} {et~al.}(2007){Durisen}, {Boss}, {Mayer}, {Nelson},
  {Quinn}, \& {Rice}}]{durisen2007}
{Durisen}, R.~H., {Boss}, A.~P., {Mayer}, L., {et~al.} 2007, Protostars and
  Planets V, 607

\bibitem[{{Eggenberger} {et~al.}(2007){Eggenberger}, {Udry}, {Chauvin},
  {Beuzit}, {Lagrange}, {S{\'e}gransan}, \& {Mayor}}]{eggenberger2007}
{Eggenberger}, A., {Udry}, S., {Chauvin}, G., {et~al.} 2007, \aap, 474, 273

\bibitem[{{Epchtein} {et~al.}(1999){Epchtein}, {Deul}, {Derriere},
  {Borsenberger}, {Egret}, {Simon}, {Alard}, {Bal{\'a}zs}, {de Batz}, {Cioni},
  {Copet}, {Dennefeld}, {Forveille}, {Fouqu{\'e}}, {Garz{\'o}n}, {Habing},
  {Holl}, {Hron}, {Kimeswenger}, {Lacombe}, {Le Bertre}, {Loup}, {Mamon},
  {Omont}, {Paturel}, {Persi}, {Robin}, {Rouan}, {Tiph{\`e}ne}, {Vauglin}, \&
  {Wagner}}]{epchtein1999}
{Epchtein}, N., {Deul}, E., {Derriere}, S., {et~al.} 1999, \aap, 349, 236

\bibitem[{{Ercolano} \& {Pascucci}(2017)}]{ercolano2017}
{Ercolano}, B., \& {Pascucci}, I. 2017, Royal Society Open Science, 4, 170114

\bibitem[{{Espaillat} {et~al.}(2007){Espaillat}, {Calvet}, {D'Alessio},
  {Bergin}, {Hartmann}, {Watson}, {Furlan}, {Najita}, {Forrest}, {McClure},
  {Sargent}, {Bohac}, \& {Harrold}}]{espaillat2007}
{Espaillat}, C., {Calvet}, N., {D'Alessio}, P., {et~al.} 2007, \apjl, 664, L111

\bibitem[{{Espaillat} {et~al.}(2017){Espaillat}, {Ribas}, {McClure},
  {Hern{\'a}ndez}, {Owen}, {Avish}, {Calvet}, \&
  {Franco-Hern{\'a}ndez}}]{espaillat2017}
{Espaillat}, C.~C., {Ribas}, {\'A}., {McClure}, M.~K., {et~al.} 2017, \apj,
  844, 60

\bibitem[{{Fang} {et~al.}(2013){Fang}, {Kim}, {van Boekel}, {Sicilia-Aguilar},
  {Henning}, \& {Flaherty}}]{fang2013}
{Fang}, M., {Kim}, J.~S., {van Boekel}, R., {et~al.} 2013, \apjs, 207, 5

\bibitem[{{Fedele} {et~al.}(2010){Fedele}, {van den Ancker}, {Henning},
  {Jayawardhana}, \& {Oliveira}}]{fedele2010}
{Fedele}, D., {van den Ancker}, M.~E., {Henning}, T., {Jayawardhana}, R., \&
  {Oliveira}, J.~M. 2010, \aap, 510, A72

\bibitem[{{Feiden}(2016)}]{feiden2016}
{Feiden}, G.~A. 2016, \aap, 593, A99

\bibitem[{{Frasca} {et~al.}(2015){Frasca}, {Biazzo}, {Lanzafame}, {Alcal{\'a}},
  {Brugaletta}, {Klutsch}, {Stelzer}, {Sacco}, {Spina}, {Jeffries}, {Montes},
  {Alfaro}, {Barentsen}, {Bonito}, {Gameiro}, {L{\'o}pez-Santiago}, {Pace},
  {Pasquini}, {Prisinzano}, {Sousa}, {Gilmore}, {Randich}, {Micela},
  {Bragaglia}, {Flaccomio}, {Bayo}, {Costado}, {Franciosini}, {Hill},
  {Hourihane}, {Jofr{\'e}}, {Lardo}, {Maiorca}, {Masseron}, {Morbidelli}, \&
  {Worley}}]{frasca2015}
{Frasca}, A., {Biazzo}, K., {Lanzafame}, A.~C., {et~al.} 2015, \aap, 575, A4

\bibitem[{{Furlan} {et~al.}(2006){Furlan}, {Hartmann}, {Calvet}, {D'Alessio},
  {Franco-Hern{\'a}ndez}, {Forrest}, {Watson}, {Uchida}, {Sargent}, {Green},
  {Keller}, \& {Herter}}]{furlan2006}
{Furlan}, E., {Hartmann}, L., {Calvet}, N., {et~al.} 2006, \apjs, 165, 568

\bibitem[{{Furlan} {et~al.}(2009){Furlan}, {Watson}, {McClure}, {Manoj},
  {Espaillat}, {D'Alessio}, {Calvet}, {Kim}, {Sargent}, {Forrest}, \&
  {Hartmann}}]{furlan2009}
{Furlan}, E., {Watson}, D.~M., {McClure}, M.~K., {et~al.} 2009, \apj, 703, 1964

\bibitem[{{Gaia Collaboration} {et~al.}(2018){Gaia Collaboration}, {Brown},
  {Vallenari}, {Prusti}, {de Bruijne}, {Babusiaux}, \&
  {Bailer-Jones}}]{gaia2018}
{Gaia Collaboration}, {Brown}, A.~G.~A., {Vallenari}, A., {et~al.} 2018, ArXiv
  e-prints, arXiv:1804.09365

\bibitem[{{Guilloteau} {et~al.}(2014){Guilloteau}, {Simon}, {Pi{\'e}tu}, {Di
  Folco}, {Dutrey}, {Prato}, \& {Chapillon}}]{guilloteau2014}
{Guilloteau}, S., {Simon}, M., {Pi{\'e}tu}, V., {et~al.} 2014, \aap, 567, A117

\bibitem[{{Haisch} {et~al.}(2001){Haisch}, {Lada}, \& {Lada}}]{haisch2001}
{Haisch}, Jr., K.~E., {Lada}, E.~A., \& {Lada}, C.~J. 2001, \apjl, 553, L153

\bibitem[{{Hardy} {et~al.}(2015){Hardy}, {Caceres}, {Schreiber}, {Cieza},
  {Alexander}, {Canovas}, {Williams}, {Wahhaj}, \& {Menard}}]{hardy2015}
{Hardy}, A., {Caceres}, C., {Schreiber}, M.~R., {et~al.} 2015, \aap, 583, A66

\bibitem[{{Harris} {et~al.}(2012){Harris}, {Andrews}, {Wilner}, \&
  {Kraus}}]{harris2012}
{Harris}, R.~J., {Andrews}, S.~M., {Wilner}, D.~J., \& {Kraus}, A.~L. 2012,
  \apj, 751, 115

\bibitem[{{Hartmann} {et~al.}(1998){Hartmann}, {Calvet}, {Gullbring}, \&
  {D'Alessio}}]{hartmann1998}
{Hartmann}, L., {Calvet}, N., {Gullbring}, E., \& {D'Alessio}, P. 1998, \apj,
  495, 385

\bibitem[{{Hauschildt} {et~al.}(1999){Hauschildt}, {Allard}, {Ferguson},
  {Baron}, \& {Alexander}}]{hauschildt1999}
{Hauschildt}, P.~H., {Allard}, F., {Ferguson}, J., {Baron}, E., \& {Alexander},
  D.~R. 1999, \apj, 525, 871

\bibitem[{{Herczeg} \& {Hillenbrand}(2008)}]{herczeg2008}
{Herczeg}, G.~J., \& {Hillenbrand}, L.~A. 2008, \apj, 681, 594

\bibitem[{{Herczeg} \& {Hillenbrand}(2014)}]{herczeg2014}
---. 2014, \apj, 786, 97

\bibitem[{{Hern{\'a}ndez} {et~al.}(2007){Hern{\'a}ndez}, {Calvet},
  {Brice{\~n}o}, {Hartmann}, {Vivas}, {Muzerolle}, {Downes}, {Allen}, \&
  {Gutermuth}}]{hernandez2007}
{Hern{\'a}ndez}, J., {Calvet}, N., {Brice{\~n}o}, C., {et~al.} 2007, \apj, 671,
  1784

\bibitem[{{Horch} {et~al.}(2014){Horch}, {Howell}, {Everett}, \&
  {Ciardi}}]{horch2014}
{Horch}, E.~P., {Howell}, S.~B., {Everett}, M.~E., \& {Ciardi}, D.~R. 2014,
  \apj, 795, 60

\bibitem[{{Jensen} {et~al.}(1994){Jensen}, {Mathieu}, \& {Fuller}}]{jensen1994}
{Jensen}, E.~L.~N., {Mathieu}, R.~D., \& {Fuller}, G.~A. 1994, \apjl, 429, L29

\bibitem[{{Jensen} {et~al.}(1996){Jensen}, {Mathieu}, \& {Fuller}}]{jensen1996}
---. 1996, \apj, 458, 312

\bibitem[{{Kelly}(2007)}]{kelly2007}
{Kelly}, B.~C. 2007, \apj, 665, 1489

\bibitem[{{Kim} {et~al.}(2009){Kim}, {Watson}, {Manoj}, {Furlan}, {Najita},
  {Forrest}, {Sargent}, {Espaillat}, {Calvet}, {Luhman}, {McClure}, {Green}, \&
  {Harrold}}]{kim2009}
{Kim}, K.~H., {Watson}, D.~M., {Manoj}, P., {et~al.} 2009, \apj, 700, 1017

\bibitem[{{Kraus} \& {Hillenbrand}(2007)}]{kraus2007}
{Kraus}, A.~L., \& {Hillenbrand}, L.~A. 2007, \apj, 662, 413

\bibitem[{{Kraus} {et~al.}(2012){Kraus}, {Ireland}, {Hillenbrand}, \&
  {Martinache}}]{kraus2012}
{Kraus}, A.~L., {Ireland}, M.~J., {Hillenbrand}, L.~A., \& {Martinache}, F.
  2012, \apj, 745, 19

\bibitem[{{Kraus} {et~al.}(2016){Kraus}, {Ireland}, {Huber}, {Mann}, \&
  {Dupuy}}]{kraus2016}
{Kraus}, A.~L., {Ireland}, M.~J., {Huber}, D., {Mann}, A.~W., \& {Dupuy}, T.~J.
  2016, \aj, 152, 8

\bibitem[{{Lada} {et~al.}(2006){Lada}, {Muench}, {Luhman}, {Allen}, {Hartmann},
  {Megeath}, {Myers}, {Fazio}, {Wood}, {Muzerolle}, {Rieke}, {Siegler}, \&
  {Young}}]{lada2006}
{Lada}, C.~J., {Muench}, A.~A., {Luhman}, K.~L., {et~al.} 2006, \aj, 131, 1574

\bibitem[{{Lafreni{\`e}re} {et~al.}(2008){Lafreni{\`e}re}, {Jayawardhana},
  {Brandeker}, {Ahmic}, \& {van Kerkwijk}}]{lafrenire2008}
{Lafreni{\`e}re}, D., {Jayawardhana}, R., {Brandeker}, A., {Ahmic}, M., \& {van
  Kerkwijk}, M.~H. 2008, \apj, 683, 844

\bibitem[{{Lissauer} \& {Stevenson}(2007)}]{lissauer2007}
{Lissauer}, J.~J., \& {Stevenson}, D.~J. 2007, Protostars and Planets V, 591

\bibitem[{{Liu} {et~al.}(2015){Liu}, {Joergens}, {Bayo}, {Nielbock}, \&
  {Wang}}]{liu2015}
{Liu}, Y., {Joergens}, V., {Bayo}, A., {Nielbock}, M., \& {Wang}, H. 2015,
  \aap, 582, A22

\bibitem[{{Lodato} {et~al.}(2017){Lodato}, {Scardoni}, {Manara}, \&
  {Testi}}]{lodato2017}
{Lodato}, G., {Scardoni}, C.~E., {Manara}, C.~F., \& {Testi}, L. 2017, \mnras,
  472, 4700

\bibitem[{{Long} {et~al.}(2017){Long}, {Herczeg}, {Pascucci}, {Drabek-Maunder},
  {Mohanty}, {Testi}, {Apai}, {Hendler}, {Henning}, {Manara}, \&
  {Mulders}}]{long2017}
{Long}, F., {Herczeg}, G.~J., {Pascucci}, I., {et~al.} 2017, \apj, 844, 99

\bibitem[{{Lopez Mart{\'{\i}}} {et~al.}(2013){Lopez Mart{\'{\i}}}, {Jimenez
  Esteban}, {Bayo}, {Barrado}, {Solano}, \& {Rodrigo}}]{lopez2013}
{Lopez Mart{\'{\i}}}, B., {Jimenez Esteban}, F., {Bayo}, A., {et~al.} 2013,
  \aap, 551, A46

\bibitem[{{Lubow} {et~al.}(2015){Lubow}, {Martin}, \& {Nixon}}]{lubow2015}
{Lubow}, S.~H., {Martin}, R.~G., \& {Nixon}, C. 2015, \apj, 800, 96

\bibitem[{{Luhman}(2004)}]{luhman2004}
{Luhman}, K.~L. 2004, \apj, 602, 816

\bibitem[{{Luhman}(2007)}]{luhman2007}
---. 2007, \apjs, 173, 104

\bibitem[{{Luhman} {et~al.}(2010){Luhman}, {Allen}, {Espaillat}, {Hartmann}, \&
  {Calvet}}]{luhman2010}
{Luhman}, K.~L., {Allen}, P.~R., {Espaillat}, C., {Hartmann}, L., \& {Calvet},
  N. 2010, \apjs, 186, 111

\bibitem[{{Luhman} {et~al.}(2008){Luhman}, {Allen}, {Allen}, {Gutermuth},
  {Hartmann}, {Mamajek}, {Megeath}, {Myers}, \& {Fazio}}]{luhman2008}
{Luhman}, K.~L., {Allen}, L.~E., {Allen}, P.~R., {et~al.} 2008, \apj, 675, 1375

\bibitem[{{Manara} {et~al.}(2016){Manara}, {Fedele}, {Herczeg}, \&
  {Teixeira}}]{manara2016}
{Manara}, C.~F., {Fedele}, D., {Herczeg}, G.~J., \& {Teixeira}, P.~S. 2016,
  \aap, 585, A136

\bibitem[{{Manara} {et~al.}(2013){Manara}, {Testi}, {Rigliaco}, {Alcal{\'a}},
  {Natta}, {Stelzer}, {Biazzo}, {Covino}, {Covino}, {Cupani}, {D'Elia}, \&
  {Randich}}]{manara2013}
{Manara}, C.~F., {Testi}, L., {Rigliaco}, E., {et~al.} 2013, \aap, 551, A107

\bibitem[{{Manara} {et~al.}(2017){Manara}, {Testi}, {Herczeg}, {Pascucci},
  {Alcal{\'a}}, {Natta}, {Antoniucci}, {Fedele}, {Mulders}, {Henning},
  {Mohanty}, {Prusti}, \& {Rigliaco}}]{manara2017}
{Manara}, C.~F., {Testi}, L., {Herczeg}, G.~J., {et~al.} 2017, \aap, 604, A127

\bibitem[{{Manoj} {et~al.}(2011){Manoj}, {Kim}, {Furlan}, {McClure}, {Luhman},
  {Watson}, {Espaillat}, {Calvet}, {Najita}, {D'Alessio}, {Adame}, {Sargent},
  {Forrest}, {Bohac}, {Green}, \& {Arnold}}]{manoj2011}
{Manoj}, P., {Kim}, K.~H., {Furlan}, E., {et~al.} 2011, \apjs, 193, 11

\bibitem[{{Matr{\`a}} {et~al.}(2012){Matr{\`a}}, {Mer{\'{\i}}n}, {Alves de
  Oliveira}, {Hu{\'e}lamo}, {K{\'o}sp{\'a}l}, {Cox}, {Ribas}, {Puga}, {Vavrek},
  {Royer}, {Prusti}, {Pilbratt}, \& {Andr{\'e}}}]{matra2012}
{Matr{\`a}}, L., {Mer{\'{\i}}n}, B., {Alves de Oliveira}, C., {et~al.} 2012,
  \aap, 548, A111

\bibitem[{{McCabe} {et~al.}(2006){McCabe}, {Ghez}, {Prato}, {Duch{\^e}ne},
  {Fisher}, \& {Telesco}}]{mccabe2006}
{McCabe}, C., {Ghez}, A.~M., {Prato}, L., {et~al.} 2006, \apj, 636, 932

\bibitem[{{Mer{\'{\i}}n} {et~al.}(2010){Mer{\'{\i}}n}, {Brown}, {Oliveira},
  {Herczeg}, {van Dishoeck}, {Bottinelli}, {Evans}, {Cieza}, {Spezzi},
  {Alcal{\'a}}, {Harvey}, {Blake}, {Bayo}, {Geers}, {Lahuis}, {Prusti},
  {Augereau}, {Olofsson}, {Walter}, \& {Chiu}}]{merin2010}
{Mer{\'{\i}}n}, B., {Brown}, J.~M., {Oliveira}, I., {et~al.} 2010, \apj, 718,
  1200

\bibitem[{{Miranda} \& {Lai}(2015)}]{miranda2015}
{Miranda}, R., \& {Lai}, D. 2015, \mnras, 452, 2396

\bibitem[{{Mo{\'o}r} {et~al.}(2017){Mo{\'o}r}, {Cur{\'e}}, {K{\'o}sp{\'a}l},
  {{\'A}brah{\'a}m}, {Csengeri}, {Eiroa}, {Gunawan}, {Henning}, {Hughes},
  {Juh{\'a}sz}, {Pawellek}, \& {Wyatt}}]{Moor2017}
{Mo{\'o}r}, A., {Cur{\'e}}, M., {K{\'o}sp{\'a}l}, {\'A}., {et~al.} 2017, \apj,
  849, 123

\bibitem[{{Mulders} \& {Dominik}(2012)}]{mulders2012}
{Mulders}, G.~D., \& {Dominik}, C. 2012, \aap, 539, A9

\bibitem[{{Mulders} {et~al.}(2017){Mulders}, {Pascucci}, {Manara}, {Testi},
  {Herczeg}, {Henning}, {Mohanty}, \& {Lodato}}]{mulders2017}
{Mulders}, G.~D., {Pascucci}, I., {Manara}, C.~F., {et~al.} 2017, \apj, 847, 31

\bibitem[{{Murphy} {et~al.}(2013){Murphy}, {Lawson}, \& {Bessell}}]{murphy2013}
{Murphy}, S.~J., {Lawson}, W.~A., \& {Bessell}, M.~S. 2013, \mnras, 435, 1325

\bibitem[{{Nguyen} {et~al.}(2012){Nguyen}, {Brandeker}, {van Kerkwijk}, \&
  {Jayawardhana}}]{nguyen2012}
{Nguyen}, D.~C., {Brandeker}, A., {van Kerkwijk}, M.~H., \& {Jayawardhana}, R.
  2012, \apj, 745, 119

\bibitem[{{Olofsson} {et~al.}(2013){Olofsson}, {Sz{\H u}cs}, {Henning}, {Linz},
  {Pascucci}, \& {Joergens}}]{olofsson2013}
{Olofsson}, J., {Sz{\H u}cs}, L., {Henning}, T., {et~al.} 2013, \aap, 560, A100

\bibitem[{{Owen} {et~al.}(2012){Owen}, {Clarke}, \& {Ercolano}}]{owen2012}
{Owen}, J.~E., {Clarke}, C.~J., \& {Ercolano}, B. 2012, \mnras, 422, 1880

\bibitem[{{Padgett} {et~al.}(2006){Padgett}, {Cieza}, {Stapelfeldt}, {Evans},
  {Koerner}, {Sargent}, {Fukagawa}, {van Dishoeck}, {Augereau}, {Allen},
  {Blake}, {Brooke}, {Chapman}, {Harvey}, {Porras}, {Lai}, {Mundy}, {Myers},
  {Spiesman}, \& {Wahhaj}}]{padgett2006}
{Padgett}, D.~L., {Cieza}, L., {Stapelfeldt}, K.~R., {et~al.} 2006, \apj, 645,
  1283

\bibitem[{{Papaloizou} \& {Pringle}(1977)}]{papaloizou1977}
{Papaloizou}, J., \& {Pringle}, J.~E. 1977, \mnras, 181, 441

\bibitem[{{Pascucci} {et~al.}(2008){Pascucci}, {Apai}, {Hardegree-Ullman},
  {Kim}, {Meyer}, \& {Bouwman}}]{pascucci2008}
{Pascucci}, I., {Apai}, D., {Hardegree-Ullman}, E.~E., {et~al.} 2008, \apj,
  673, 477

\bibitem[{{Pascucci} {et~al.}(2009){Pascucci}, {Apai}, {Luhman}, {Henning},
  {Bouwman}, {Meyer}, {Lahuis}, \& {Natta}}]{pascucci2009}
{Pascucci}, I., {Apai}, D., {Luhman}, K., {et~al.} 2009, \apj, 696, 143

\bibitem[{{Pascucci} {et~al.}(2016){Pascucci}, {Testi}, {Herczeg}, {Long},
  {Manara}, {Hendler}, {Mulders}, {Krijt}, {Ciesla}, {Henning}, {Mohanty},
  {Drabek-Maunder}, {Apai}, {Sz{\H u}cs}, {Sacco}, \&
  {Olofsson}}]{pascucci2016}
{Pascucci}, I., {Testi}, L., {Herczeg}, G.~J., {et~al.} 2016, \apj, 831, 125

\bibitem[{{Pinilla} {et~al.}(2015){Pinilla}, {van der Marel}, {P{\'e}rez}, {van
  Dishoeck}, {Andrews}, {Birnstiel}, {Herczeg}, {Pontoppidan}, \& {van
  Kempen}}]{pinilla2015}
{Pinilla}, P., {van der Marel}, N., {P{\'e}rez}, L.~M., {et~al.} 2015, \aap,
  584, A16

\bibitem[{{Reipurth} \& {Zinnecker}(1993)}]{reipurth1993}
{Reipurth}, B., \& {Zinnecker}, H. 1993, \aap, 278, 81

\bibitem[{{Ribas} {et~al.}(2017){Ribas}, {Espaillat}, {Mac{\'{\i}}as}, {Bouy},
  {Andrews}, {Calvet}, {Naylor}, {Riviere-Marichalar}, {van der Wiel}, \&
  {Wilner}}]{ribas2017}
{Ribas}, {\'A}., {Espaillat}, C.~C., {Mac{\'{\i}}as}, E., {et~al.} 2017, \apj,
  849, 63

\bibitem[{{Rosotti} \& {Clarke}(2018)}]{rosotti2018}
{Rosotti}, G.~P., \& {Clarke}, C.~J. 2018, \mnras, 473, 5630

\bibitem[{{Sacco} {et~al.}(2017){Sacco}, {Spina}, {Randich}, {Palla}, {Parker},
  {Jeffries}, {Jackson}, {Meyer}, {Mapelli}, {Lanzafame}, {Bonito}, {Damiani},
  {Franciosini}, {Frasca}, {Klutsch}, {Prisinzano}, {Tognelli},
  {Degl'Innocenti}, {Prada Moroni}, {Alfaro}, {Micela}, {Prusti}, {Barrado},
  {Biazzo}, {Bouy}, {Bravi}, {Lopez-Santiago}, {Wright}, {Bayo}, {Gilmore},
  {Bragaglia}, {Flaccomio}, {Koposov}, {Pancino}, {Casey}, {Costado}, {Donati},
  {Hourihane}, {Jofr{\'e}}, {Lardo}, {Lewis}, {Magrini}, {Monaco},
  {Morbidelli}, {Sousa}, {Worley}, \& {Zaggia}}]{sacco2017}
{Sacco}, G.~G., {Spina}, L., {Randich}, S., {et~al.} 2017, \aap, 601, A97

\bibitem[{{Simon} {et~al.}(2017){Simon}, {Guilloteau}, {Di Folco}, {Dutrey},
  {Grosso}, {Pi{\'e}tu}, {Chapillon}, {Prato}, {Schaefer}, {Rice}, \&
  {Boehler}}]{simon2017}
{Simon}, M., {Guilloteau}, S., {Di Folco}, E., {et~al.} 2017, \apj, 844, 158

\bibitem[{{Skrutskie} {et~al.}(2006){Skrutskie}, {Cutri}, {Stiening},
  {Weinberg}, {Schneider}, {Carpenter}, {Beichman}, {Capps}, {Chester},
  {Elias}, {Huchra}, {Liebert}, {Lonsdale}, {Monet}, {Price}, {Seitzer},
  {Jarrett}, {Kirkpatrick}, {Gizis}, {Howard}, {Evans}, {Fowler}, {Fullmer},
  {Hurt}, {Light}, {Kopan}, {Marsh}, {McCallon}, {Tam}, {Van Dyk}, \&
  {Wheelock}}]{skrutskie2006}
{Skrutskie}, M.~F., {Cutri}, R.~M., {Stiening}, R., {et~al.} 2006, \aj, 131,
  1163

\bibitem[{{Stassun} {et~al.}(2014){Stassun}, {Feiden}, \&
  {Torres}}]{stassun2014}
{Stassun}, K.~G., {Feiden}, G.~A., \& {Torres}, G. 2014, \nar, 60, 1

\bibitem[{{Sz{\H u}cs} {et~al.}(2010){Sz{\H u}cs}, {Apai}, {Pascucci}, \&
  {Dullemond}}]{szucs2010}
{Sz{\H u}cs}, L., {Apai}, D., {Pascucci}, I., \& {Dullemond}, C.~P. 2010, \apj,
  720, 1668

\bibitem[{{Venuti} {et~al.}(2015){Venuti}, {Bouvier}, {Irwin}, {Stauffer},
  {Hillenbrand}, {Rebull}, {Cody}, {Alencar}, {Micela}, {Flaccomio}, \&
  {Peres}}]{venuti2015}
{Venuti}, L., {Bouvier}, J., {Irwin}, J., {et~al.} 2015, \aap, 581, A66

\bibitem[{{Voirin} {et~al.}(2017){Voirin}, {Manara}, \& {Prusti}}]{voirin2017}
{Voirin}, J., {Manara}, C.~F., \& {Prusti}, T. 2017, ArXiv e-prints,
  arXiv:1710.04528

\bibitem[{{Wahhaj} {et~al.}(2010){Wahhaj}, {Cieza}, {Koerner}, {Stapelfeldt},
  {Padgett}, {Case}, {Keller}, {Mer{\'{\i}}n}, {Evans}, {Harvey}, {Sargent},
  {van Dishoeck}, {Allen}, {Blake}, {Brooke}, {Chapman}, {Mundy}, \&
  {Myers}}]{wahhaj2010}
{Wahhaj}, Z., {Cieza}, L., {Koerner}, D.~W., {et~al.} 2010, \apj, 724, 835

\bibitem[{{Wang} {et~al.}(2014){Wang}, {Xie}, {Barclay}, \&
  {Fischer}}]{wang2014a}
{Wang}, J., {Xie}, J.-W., {Barclay}, T., \& {Fischer}, D.~A. 2014, \apj, 783, 4

\bibitem[{{Weidenschilling}(1977)}]{weidenschilling1977}
{Weidenschilling}, S.~J. 1977, \mnras, 180, 57

\bibitem[{{Yen} {et~al.}(2018){Yen}, {Koch}, {Manara}, {Miotello}, \&
  {Testi}}]{yen2018}
{Yen}, H.-W., {Koch}, P.~M., {Manara}, C.~F., {Miotello}, A., \& {Testi}, L.
  2018, ArXiv e-prints, arXiv:1804.06272

\bibitem[{{Zhu} {et~al.}(2012){Zhu}, {Nelson}, {Dong}, {Espaillat}, \&
  {Hartmann}}]{zhu2012}
{Zhu}, Z., {Nelson}, R.~P., {Dong}, R., {Espaillat}, C., \& {Hartmann}, L.
  2012, \apj, 755, 6

\end{thebibliography}


\clearpage
\appendix
\section{Descriptions of individual faint disks} \label{sec:faint-detail}

\subsection{CHXR71}
CHXR71 is an M3 star \citep{luhman2004,manara2016} with an M5 secondary located at a projected separation of $0.\arcsec56$, or 100 AU \citep{lafrenire2008,daemgen2013}. 
In the optical spectrum, which is dominated by the primary component, the narrow and weak H$\alpha$ line emission is consistent with chromospheric emission \citep{luhman2004,nguyen2012,frasca2015}.  Fits to the UV continuum emission led to an accretion rate that is consistent with chromospheric activity \citep{manara2017}. 
The warm dust near one of the stars produces at most a marginal excess at $\la$8 $\mu$m and a small excess at longer wavelengths, indicating substantial dust clearing within the innermost few AU of the disk.

\subsection{CHXR30A}
CHXR30A is a K7--K8 star \citep{luhman2004,manara2016} with a companion with a \textit{K}$_s$-band contrast of 0.85 at a projected separation of 0.$\arcsec$46, or 83 AU  \citep{lafrenire2008}.  The \textit{K}$_s$ band contrast leads to an estimated mass ratio of $\sim$0.4--0.5 in the \citet{baraffe2015} models of pre-main sequence evolution, assuming that neither component has near-IR emission affected by a disk.  Any accretion onto the primary is either very weak or undetected, with line emission that is consistent with chromospheric emission \citep{luhman2004,manara2017}.

The SED of CHXR30A has a steep $n_{2-13}$ index and a significant flux deficit at wavelengths $\la$8 $\mu$m, indicating that dust grains in the inner disk are largely depleted \citep{manoj2011}. 
The continuum excess longward of 13 $\mu$m is much fainter than that of typical Class II disks, though silicate features at 10 $\mu$m and 20 $\mu$m have moderate strengths that are comparable to Class II disks. 
The decreasing trend in flux towards longer wavelengths in the Spitzer-IRS spectrum (13--31 $\mu$m, \citealt{manoj2011}) differs significantly from the opposite trend commonly seen in transition disks \citep{calvet2005,espaillat2007}, which means that any inner dust wall is not prominent.

\subsection{T54}
T54 is a K0 star \citep{manara2016} with a K7 secondary at a projected separation of 0.$\arcsec$24, or 43 AU \citep{daemgen2013}.  Based on a weak UV continuum excess, \citet{manara2016} reported a low accretion luminosity, comparable to chromospheric noise.
\citet{espaillat2017} carefully investigated accretion properties of this target from H$_2$, Br$\gamma$, \ion{He}{1}, and H$\alpha$ line emission and concluded that it is at most barely accreting.

\citet{furlan2009} and \citet{kim2009} classified  T54 as a transition disk candidate because the IR SED is photospheric at wavelengths $<$10 $\mu$m and rises longward of 15 $\mu$m. The 10 $\mu$m silicate feature is below the detection limits but PAH emission at 11.3 $\mu$m is prominent \citep{espaillat2017}. Far-IR excess ($\lambda$$>$70 $\mu$m) from \textit{Herschel} photometry is largely contaminated by a nearby unidentified source \citep{matra2012}. Based on low dust mass and gas detection in the inner disk, \citet{espaillat2017} argue that T54 is a young debris disk.

\subsection{CHXR20}
CHXR20 is a K6 star \citep{luhman2004,manara2017} with a wide companion separated by 28.$\arcsec$46 \citep{kraus2007} and no known close companion from either high-resolution spectroscopy \citep{nguyen2012} or Adaptive Optics imaging \citep{lafrenire2008}. The star is weakly accreting, at a rate of $1.95\times10^{-9}$\textit{M}$_{\sun}$ yr$^{-1}$ as measured from the UV excess \citep{manara2017}.  Weak accretion is consistent with past measurements of H$\alpha$ emission, which has an equivalent width consistent with a chromsopheric origin \citep{luhman2004,frasca2015} but is also broad \citep{sacco2017} and includes absorption components in at least some epochs.
 The SED of CHXR20 shows a deficit of near-IR excess shortward of 6 $\mu$m, similar to transitional disks, while the spectral slope at longer wavelength does not rise, as expected if the disk had a puffed-up wall typical of many transition disks \citep{espaillat2007}.  The mid-IR excess is fainter than typical Class II disks but is slightly brighter than that for CHXR71 and CHXR30A.

\subsection{ISO91}
\begin{figure*}[h]
\centering
    \includegraphics[width=0.9\textwidth]{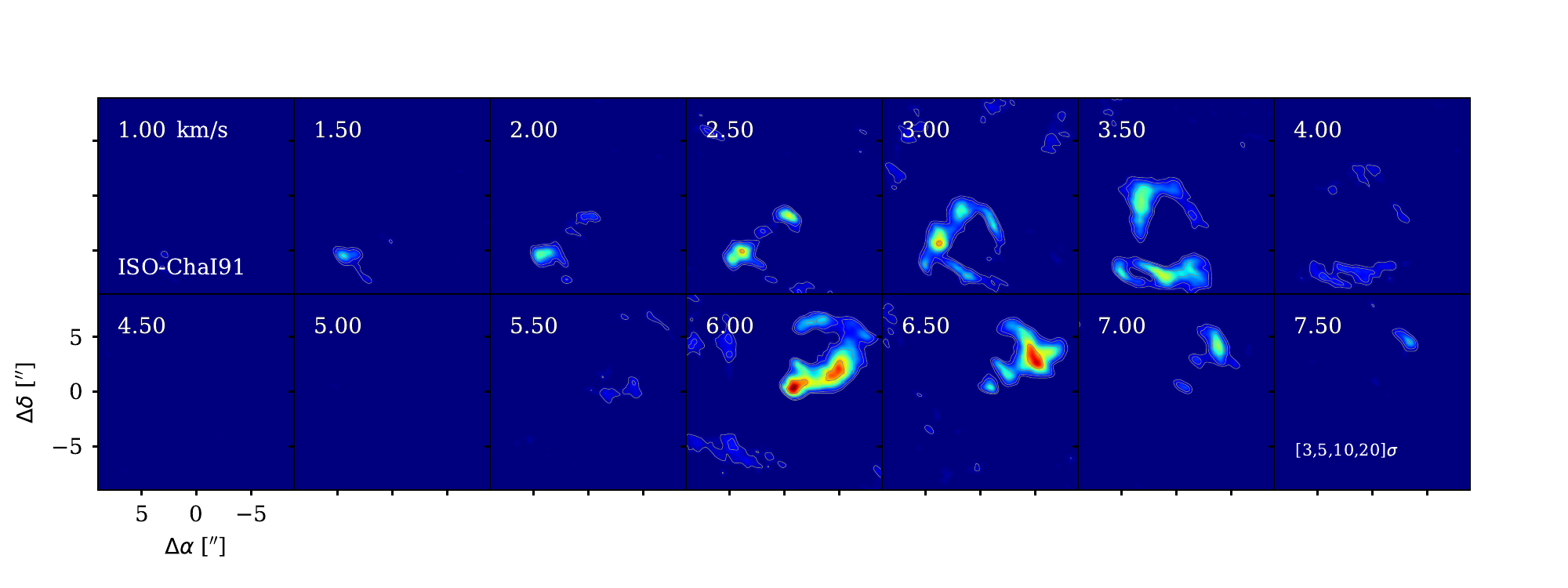}
    \caption{The $^{12}$CO J = 3-2 channel maps for ISO91, with channel velocity shown in the upper left corner. Large scale CO outflow extends to $\sim$10$\arcsec$.  \label{fig:12co}}
\end{figure*}

ISO91 is an M3 star with high extinction of \textit{A$_J$} = 4.51, unlike most known members in Cha I \citep{luhman2007}. Due to its high extinction, information on multiplicity is not available. Accretion diagnostics from H$\alpha$ or other measurements are also not available \citep{manara2017}.
After correction for extinction, the IR SED is similar to typical Class II disks, though with average lower IR excess.  The 10 $\mu$m and 20 $\mu$m  silicate emission is stronger than most other disks and than prediction from typical disk models \citep{manoj2011}. Though continuum emission is only detected at 3.8$\sigma$ at source center, a large-scale $^{12}$CO outflow (see Figure \ref{fig:12co}) extends to $\sim$10$\arcsec$,  with centroid velocity consistent with the average system velocity for Chamaeleon I sample \citep{nguyen2012}. The SED and mid-IR spectral slope appear disk-like \citep{dunham2015} and the source is not clearly detected with Herschel/PACS \citep{ribas2017}.  The outflow and extinction both indicate that ISO91 may still be embedded in an envelope and at an earlier stage of evolution than the other sources in this sample.  Single-dish observations would help to reveal the presence of any envelope. 

\subsection{T51}
T51 is a K2 star with an M2 secondary \citep{manara2016} located at a projected separation of  $1.\arcsec9$ \citep{reipurth1993,correia2006,lafrenire2008}. 
The binary was resolved in VLT/X-shooter observations, with measured accretion rates of both components comparable to typical Class II disks for each relevant spectral type \citep{manara2016}.  Most spectra of H$\alpha$ confirm the presence of accretion, with clear accretion signatures in line strength and line width \citep{wahhaj2010,frasca2015}.  An initial non-detection of H$\alpha$ emission by \citet{covino1997} may indicate that the presence of accretion is variable.  The near-IR excess of T51 indicates an optically thick inner disk, again similar to typical Class II disks. However, the 13--31 $\mu$m spectral index is particularly weak, which \citet{furlan2009} and \citet{manoj2011} explain as an outward truncation by the $\sim$2$\arcsec$ companion.

\section{SEDs for brighter sources}

\begin{figure*}[h]
\centering
    \includegraphics[width=0.95\textwidth, trim=0 560 0 10]{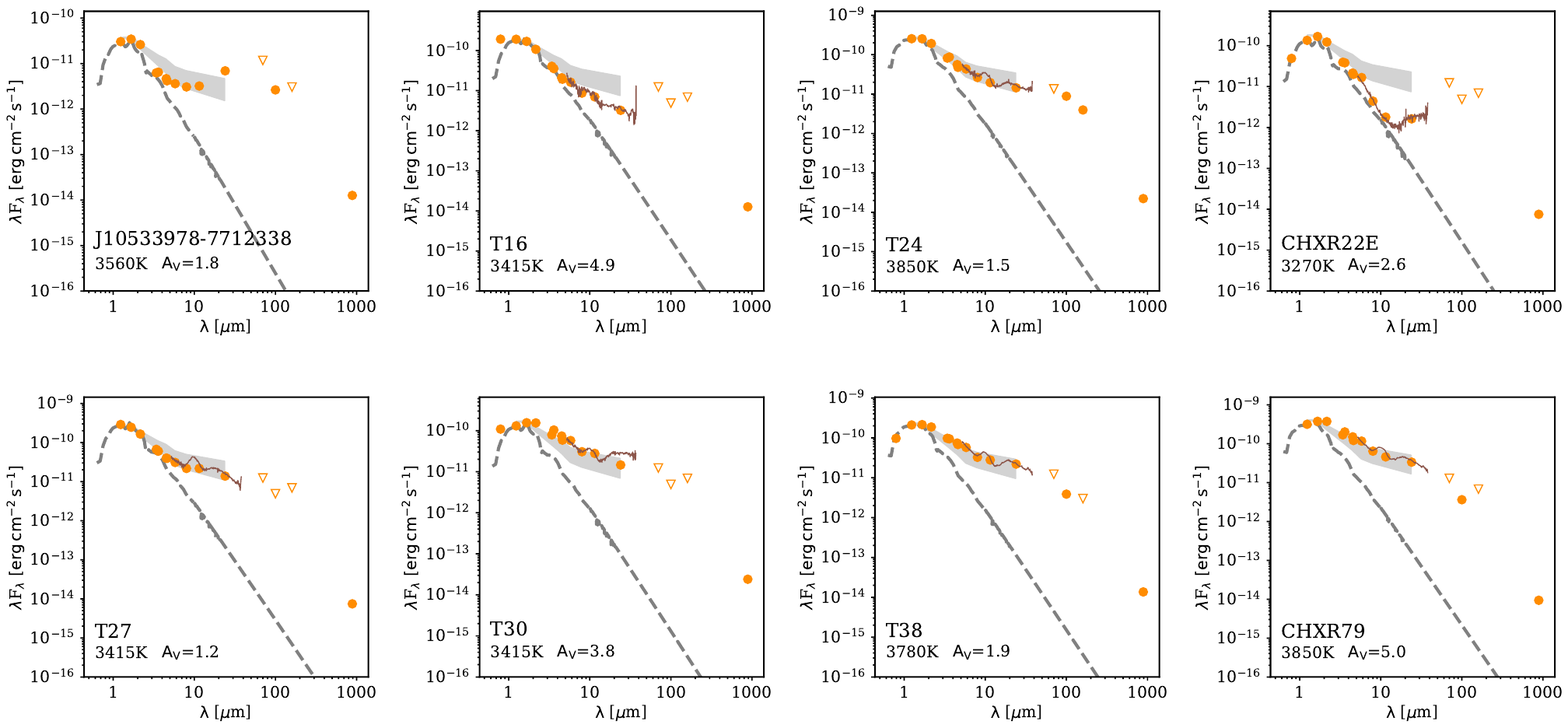}
    \caption{SEDs for the brighter sources in our ALMA Cycle 3 observation. Same as Figure \ref{fig:sed_non}. \label{fig:sed_bright}}
\end{figure*}

\end{CJK*}
\end{document}